\documentclass[%
 reprint,
superscriptaddress,
 amsmath,amssymb,
 noeprint,
 aps,
 prx,
]{revtex4-2}
\usepackage[margin=1in]{geometry}
\usepackage{graphicx} 
\usepackage[colorlinks=true,linkcolor=blue,allcolors=blue]{hyperref}
\usepackage{bm,amsmath}
\usepackage[capitalize,nameinlink]{cleveref}
\usepackage{amsthm}
\usepackage{enumitem}
\usepackage{amssymb}
\usepackage{siunitx}
 \usepackage{mathtools}
\usepackage{microtype}

\renewcommand{\vec}{\bm}
\newcommand{\m}{m} 
\newcommand{\comment}[1]{}

\def\equationautorefname~#1\null{Eq. (#1)\null}

\newcommand{\appref}[1]{\hyperref[#1]{App.~\ref*{#1}}}

\usepackage{bbm}

\usepackage{amsthm}
\usepackage{tabularx} 
\theoremstyle{definition}

\newcommand {\bx}{\mathbf{x}}
\newcommand {\bz}{\mathbf{z}}
\newcommand {\cF}{{\cal F}}
\newcommand{\bea}{\begin{eqnarray}}
\newcommand{\eea}{\end{eqnarray}}
\newcommand{\beaa}{\begin{eqnarray*}}
\newcommand{\eeaa}{\end{eqnarray*}}
\usepackage{makecell}

\begin{document}

\title{Contracting Tensor Networks with Generalized Belief Propagation}

\author{Joseph Tindall}
\affiliation{%
Center for Computational Quantum Physics, Flatiron Institute, New York, NY, 10010, USA
}%

\author{Grace M. Sommers}
\affiliation{%
Center for Computational Quantum Physics, Flatiron Institute, New York, NY, 10010, USA
}%
\affiliation{Physics Department, Princeton University, Princeton, NJ 08544, USA}

\author{Hilbert Kappen}
\affiliation{%
Center for Computational Quantum Physics, Flatiron Institute, New York, NY, 10010, USA
}%
\affiliation{Donders Institute, Radboud University, Nijmegen, the Netherlands}

\date{\today}

\begin{abstract}
Recent years have seen a growing interest in the use of belief propagation (BP), an algorithm originally introduced for performing {\color{black}classical} statistical inference on graphical models, for approximate, but highly efficient tensor network contraction. {\color{black}BP is a corner case of generalized belief propagation (GBP), a family of classical inference methods in which messages are passed within a hierarchy of overlapping regions. Here, we detail how to apply GBP to tensor network contraction. We implement GBP with different region choices on finite and infinite tensor networks in two and three dimensions, solving the corresponding fixed point equations numerically and, in tractable cases, analytically.} {\color{black}We consider a range of quantum and classical examples, including frustrated Ising models, three-dimensional ice models, independent-set counting on irregular graphs, and the norms of the deformed AKLT state and random tensor network states. Across our examples, including the quantum ones, converged GBP often reduces errors by several orders of magnitude compared with BP.}
\end{abstract}

\maketitle

\section{Introduction}
Tensor networks are a key tool for representing and manipulating structured, correlated data with large numbers of degrees of freedom. Partition functions, area-law quantum states and smooth mathematical functions on exponentially fine grids are all examples of objects that can be compactly described with a tensor network of appropriate topology \cite{White1992, Verstraete2006, Oseledets2011, Kourtis2019, Lubasch2018, Guarianov2022}. The presence of loops in the tensor network, however, can make contraction --- which is required for extracting physically relevant information --- a very computationally demanding task.
\par A large number of algorithms have been developed to tackle this problem. These typically make approximations in order to render the problem tractable within a computational time that scales polynomially with the size of the network. Notable examples include the Boundary Matrix Product State algorithm for open boundary planar topologies \cite{Murg2007, Lubasch2014}, the Corner Transfer Matrix Renormalization Group algorithm for infinite planar topologies \cite{Nishino1996, Orus2009} and index slicing and compression-based approaches for more general topologies \cite{Gray2021, Pan2022, Gray2024}.

\par The last five years --- thanks to an understanding of the relationship between tensor networks and factor graphs \cite{robeva2017dualitygraphicalmodelstensor} --- have seen a growing interest in, and development of, belief propagation (BP) and loop/cluster-based corrections {\color{black}around the BP fixed point} for the problem of tensor network contraction \cite{Alkabetz2021, tindall2023gauging, evenbly2025loopseriesexpansionstensor, gray2025tensornetworkloopcluster, midha2025beyond, PhysRevB.108.125111, midha2026beliefpropagationtensornetwork}. BP is a well-established algorithm for performing statistical inference on factor graphs, having been successfully applied to problems in error correction \cite{frey1998revolution}, compressed sensing \cite{krzakala2012statistical} and statistical physics \cite{bethe1935statistical,mezard2006reconstruction}. In the context of tensor networks, BP offers a highly efficient, but typically less controlled, path to contraction than the aforementioned methods.
It has successfully been {\color{black}incorporated as a contraction backend within larger tensor-network based workflows to enable} a number of state-of-the-art, large-scale two and three-dimensional quantum dynamics simulations with tensor networks \cite{tindall2025dynamicsdisorderedquantumsystems, Tindall2023, chan2024, Park_2025, vovrosh2025simulatingdynamicstwodimensionaltransversefield} --- with simulations of a comparable accuracy not being possible with more controlled contraction algorithms due to their significant computational cost.

\par 
{\color{black}An appealing part of the BP algorithm is that it has a corresponding variational formulation: its fixed points correspond to stationary points of an approximate ``Bethe'' free energy~\cite{NIPS2000_61b1fb3f}.} The approximation of the free energy is done by replacing the joint distribution on all variables by a set of distributions on subsets of variables (indices in the language of tensor networks) and their intersections (known as regions).
The free energy formulation of BP then allows for generalization by considering extended, overlapping regions of the factor graph. This idea was first developed by Kikuchi as the Cluster Variation Method in the context of the cubic Ising model \cite{kikuchi51}.
A hierarchy of messages between these regions is then used to approximate the marginals of the underlying distribution by minimizing the so-called Kikuchi free energy. The free energy formulation of BP and its generalization (GBP) was introduced into the machine learning community in Ref.~\cite{NIPS2000_61b1fb3f}, establishing the connection to earlier message passing methods in graphical Bayesian models \cite{pearl88,murphy99}. In general, GBP can capture more complex loop correlations than BP (which we refer to as ``simple BP'' from hereon).

\par In this work we demonstrate how to apply GBP to the problem of tensor network contraction. We derive relevant update equations for the GBP message tensors and demonstrate how to use these to both evaluate the Kikuchi free energy and approximate derivatives of the tensor network --- which are necessary for measuring observables and optimizing local tensors. The simple BP algorithm is then recovered as a corner case of GBP by an appropriate choice of initial regions. In cases where the GBP algorithm converges we find it yields stationary solutions which are much more accurate than simple BP. This is especially true in the presence of frustration, where large correlations and small loops lead BP, and thus loop-based corrections around the BP fixed point, astray. Meanwhile, when the tensor network has a significant fraction of negative or complex numbers present we find significant convergence issues and relate this to the fact the GBP update equations do not generally preserve positive semi-definiteness of the message tensors, despite the local factors in the tensor network possessing this property.

As numerical and analytical examples, we apply the GBP algorithm to approximately contract a number of finite and infinite tensor networks in both two and three dimensions. These examples represent classical partition functions, counting problems and quantum wavefunctions. Specifically, we study the following:
\begin{itemize}
{\color{black}\item The fully frustrated Ising model on a square lattice and the Ising antiferromagnet on the triangular lattice. For these models, GBP gives accurate low-temperature estimates of thermodynamic quantities while BP and loop corrections yield poor results, including an unphysical negative entropy.}
\item The ground state degeneracy of three-dimensional ice-type models, where, upon taking the fundamental voxels of the tensor network as regions, GBP is able to significantly improve upon BP and obtain values within $0.06 \%$ of current best estimates \cite{Kolafa2014, Hayashi_2021, CTMRGIce1} using a sparse-tensor implementation.
{\color{black}\item Counting independent sets on disordered planar graphs, where GBP improves on BP by roughly three orders of magnitude and is easily scalable to thousands of sites with an error that doesn't increase with system size.}
{\color{black}\item The deformed AKLT state on the honeycomb lattice, where GBP improves the free energy, reduces the spurious transverse anisotropy at small $a$, and locates the second-order AKLT-N\'eel transition closer to the true critical point than simple BP.}
\item The norm of randomly generated tensor network states {\color{black}on square and cubic lattices} with variable fraction of negative entries in the individual tensors. {\color{black}For small fractions of negative entries, GBP converges reliably, but convergence drops sharply as this fraction grows. This loss of convergence} coincides with a large increase in the errors of other algorithms such as BP, cluster corrections and boundary MPS with fixed dimension, {\color{black}suggestive of a hardness transition in the thermodynamic limit.}
\end{itemize}

\section{Preliminaries}

\begin{figure*}[t]
    \centering
    \includegraphics[width = \textwidth]{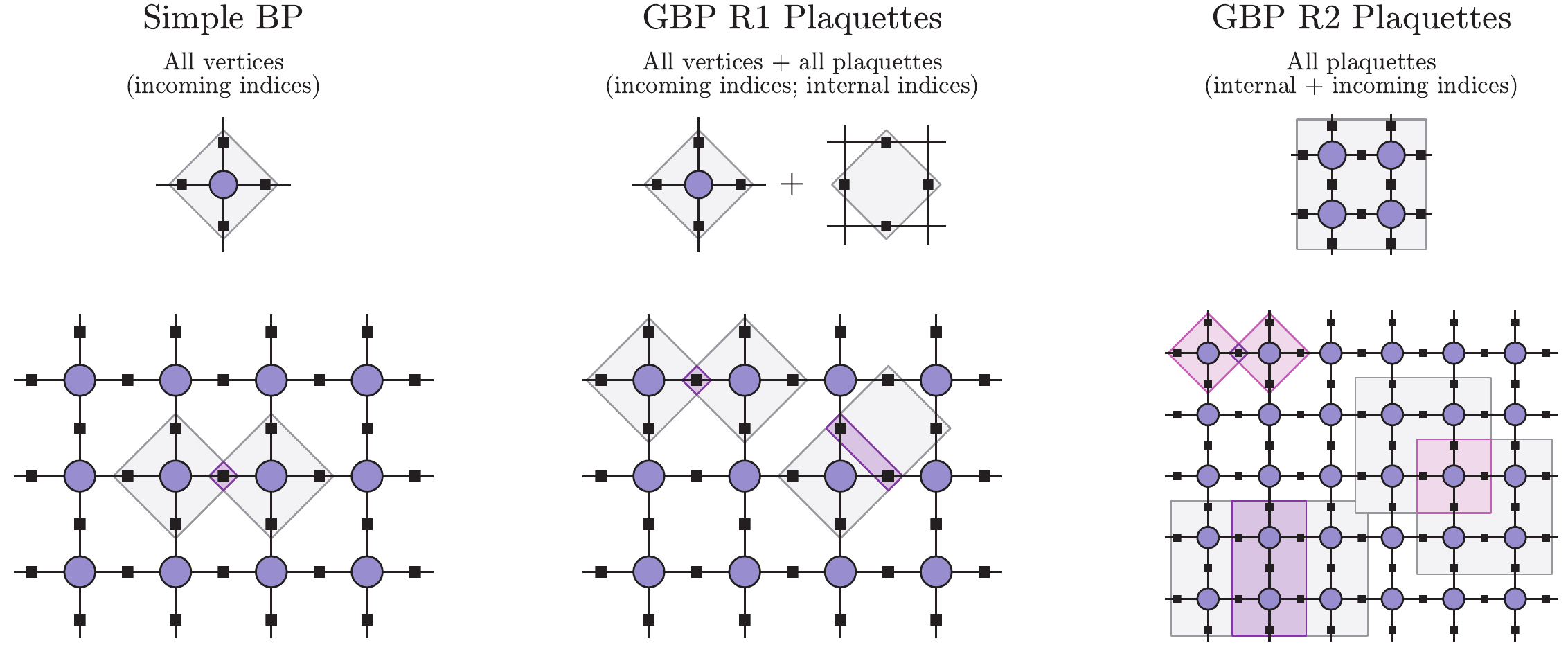}
    \caption{Examples of some of the GBP parent region choices (and resulting intersections) used in this work for the case of a square lattice tensor network. Tensors are denoted as {\color{black}circles} and indices have been made explicit as black edges with a square in the centre. {\color{black}Parent regions are outlined in gray, while child regions are shaded in purple and pink.} The intersection of parent regions populates the initial child regions. Further intersections of children (where necessary) further populate the child regions. The message tensors live in these intersections and possess the corresponding indices.
    \textbf{Simple BP)} There is a parent region for the indices associated to each and every tensor in the network. \textbf{GBP R1 Plaquettes)} There is a parent region for the indices associated to each tensor in the network and a parent region for the internal indices in each of the fundamental plaquettes of the network. \textbf{GBP R2 Plaquettes)} There is a parent region associated with each of the fundamental plaquettes in the network, where all the associated indices (both internal and incoming from the rest of the network) are included.}
    \label{fig:TN}
\end{figure*}

\subsection{Tensor Networks} 
A tensor is a function $t_{v}(\mathbf{x}_{v}) = t_{v}(x_{1}, x_{2}, x_{3} \hdots, x_{n_v})$  mapping discrete-valued variables $\mathbf{x}_{v} = \{x_{1}, x_{2}, x_{3} \hdots, x_{n_v} \}$, which we refer to as its indices, to a complex number. 
A tensor network $Z$ is a network of tensors $\mathbf{t} = \{t_{1}, t_{2}, \hdots t_{N} \}$ where each vertex represents a tensor and
an edge $t_{v} \Leftrightarrow t_{v'}$ is present if, and only if, $t_{v}$ and $t_{v'}$ share a common index/ variable. In this work we will assume no index appears more than twice in the network - and thus each index is either common to two tensors or unique to one tensor in the network. We define $\mathbf{x} = \cup_{v=1}^{N}\mathbf{x}_{v}$ as the set of all indices in the network and $\tilde{\mathbf{x}} \subseteq \mathbf{x}$ as the set of all ``common" indices in the network (i.e. those that appear in two tensors). 

In a tensor network, summation is implied over this set of common indices $\tilde{\mathbf{x}}$, i.e.
\begin{equation}
Z(\mathbf{x}  \setminus \tilde{\mathbf{x}}) = \sum_{\tilde{\mathbf{x}}}\prod_{v=1}^{N}t_{v}(\mathbf{x}_v) 
\label{Eq:TensorNetwork}
\end{equation} 
expressing the fact the network represents a graphical decomposition of a single tensor $Z(\mathbf{x}  \setminus \tilde{\mathbf{x}})$. We emphasize that the product in Eq.~(\ref{Eq:TensorNetwork}) is an element-wise (Hadamard) product over tensors. We will make it explicit in the case that a product should be interpreted as tensor contraction instead.

There is a simple, direct relationship between tensor networks and factor graphs \cite{Pancotti2023, Alkabetz2021}. Specifically, the function $ T(\mathbf{x})  = \prod_{v=1}^{N}t_{v}(\mathbf{x}_v)$ in Eq.~\ref{Eq:TensorNetwork} is in factorized form and so defines a factor graph where the tensors $t_v, v=1,\ldots,N$ correspond to the factor nodes and the indices $x=(x_1,\ldots,x_n)$ are the variable nodes. An edge between factor node $t_v$ and variable node $x_i$ exists if $x_i$ is one of the indices of $t_v$.

\subsection{Norm networks}
Frequently, in the context of many-body physics, a tensor network is used to represent a quantum many body state $\vert \psi \rangle=\sum_s \psi(s) \vert s \rangle$ with $s=s_1,\ldots,s_N$. The physical degrees of freedom are encoded in the indices $s_{v}$, one for each tensor, as $\psi(s) = \sum_{\mathbf {z}} \prod_{v=1}^N \psi_v(\mathbf{z}_v,s_v)$  and the common indices $\tilde{\mathbf{z}}$ between these tensors mediate entanglement and correlations. A typical objective is to contract, or partially contract, the norm tensor network $Z = \langle \psi \vert \psi \rangle=\sum_{\mathbf{x}}\prod_{v=1}^N t_v(\mathbf{x}_v)$ made up of local ``double factors" $t_{v}(\mathbf{x}_{v}) = t_{v}(\mathbf{z}_{v}, \mathbf{z}'_{v}) = \sum_{s_{v}}\psi_{v}(\mathbf{z}_{v}, s_{v})\psi^{*}_{v}(\mathbf{z}'_{v}, s_{v})$ formed from tracing out the physical degrees of freedom. Such tensor networks are commonly called norm networks or ``double edge" factor graphs \cite{Cao_2017} where two indices (the corresponding bra and ket index) live on each edge. The double factors can contain negative or complex entries but possess the important property that they are positive semi-definite when interpreted as a map from the bra to the ket virtual degrees of freedom. 
\subsection{Scalar tensor networks and free energy}
A norm network is one example of a scalar tensor network, i.e., a tensor network in which $Z$ is a scalar and so $\tilde{\bx} = \bx$ in ~\autoref{Eq:TensorNetwork}. By analogy to classical statistical mechanics, we will call $Z$ the partition function, and  $-\log Z$ the free energy. The GBP approximation, which we introduce below, is based on the relation between free energies and probability distributions. For non-negative $T$ with $Z>0$, the distribution of the form $p(\bx)=\frac{1}{Z}T(\bx)$ with $T(\bx)=\prod_v t_v(\bx_v)$ and $Z = \sum_{\bx} T(\bx)$ minimizes the free energy
\begin{equation}
\cF(p)=\sum_\bx p(\bx) \log \frac{p(\bx)}{T(\bx)}
\label{eq:F}
\end{equation}
under the constraint of normalization $\sum_\bx p(\bx)=1$. \autoref{eq:F}
is the difference of the expected energy $-\sum_{\bx}p(\bx)\log T(\bx)$ and the entropy $-\sum_{\bx}p(\bx)\log p(\bx)$. The minimum of $\cF$ is $\min_{p} \cF(p)=-\log Z$. 

For the norm network, $Z$  
resembles the partition sum of a probability distribution, with the important difference that the $t_v$ are positive semi-definite (psd) Hermitian matrices instead of positive scalars as is usually the case. In particular, $t_v(\bx_v)$ can be negative or complex, a point to which we will return.

\section{Generalized Belief Propagation for Tensor Networks}
For large state spaces, the sum over $\bx$ in~\autoref{eq:F} is intractable and thus so is obtaining marginals or evaluating $Z$. This motivates the use of the (generalized) belief propagation algorithm. In this section we derive (one variant of) the GBP algorithm, then discuss how simple BP and block BP \cite{PhysRevB.108.125111} are obtained as special cases and comment upon the limitations and computational complexity of the algorithm.

\begin{figure*}[t]
    \centering
    \includegraphics[width = \textwidth]{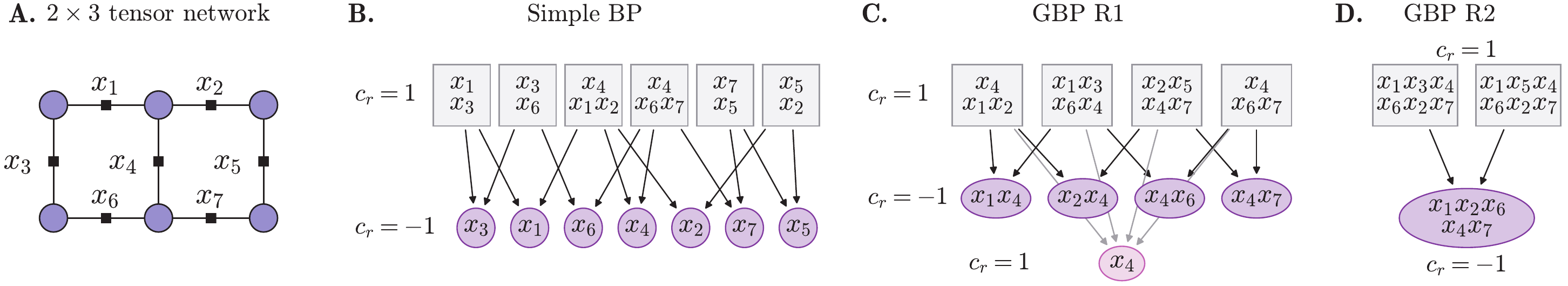}
    \caption{\textbf{A.} $2 \times 3$ tensor network with all indices $x_{i}$ (black squares) labelled. \textbf{B-D.} Region diagrams for some of the parent region choices considered in this work. Parent regions are shown at the top in rectangular boxes and always have counting number $c_{r} = 1$. Repeated intersections lead to  series of child regions (elliptical boxes) with the given counting numbers. There is a direct relationship between parent and child if that child is a subset of the parent region.
    GBP message tensors $\m_{ab}(\mathbf{x}_{b})$ live between parents and children and serve to define the marginals $p_{r}(\mathbf{x}_{r})$ for each region in the diagram. These marginals approximate the contraction of the tensor network over all indices excluding those in the region.}
    \label{fig:GBPRegions}
\end{figure*}

\subsection{Region-based approximations}
At the heart of (G)BP algorithm is the idea that we can make the sum over configurations $\bx$ in~\autoref{eq:F} tractable by approximating $p(\bx)$ in terms of a set of {\em beliefs }$p_r(\bx_r)$. Here $\bx_r$ is a subset of the indices of the tensor network $\bx$, which defines  a ``region'' $r$. These beliefs represent our approximation for the  marginalization of $p(\bx)$ onto the given region, i.e. $p_r(\bx_r) \approx \sum_{\mathbf{x} \setminus \mathbf{x}_{r}}p(\bx)$. 
 
The regions are determined by first choosing a set $P$ of ``parent" regions of the tensor network.
The only condition on the set $P$ is that for each tensor $t_{v}(\mathbf{x}_{v})$ in the network there exists at least one member $a \in P$ such that $\bx_v\subseteq \bx_a$. Note that the tensors $t_v$ may share indices, and so the parent regions will, in general, overlap.  The choice of parent regions is a crucial part of the GBP approach which determines its accuracy, convergence properties and computational complexity.

Starting from a chosen set of parent regions, we now identify a set of child regions $C$. The set $C$ is populated recursively. First, we populate it with the intersection of all unique pairs $(a, a')$ of parent regions. The region $b=a\cap a'$ contains the indices $\bx_b=\bx_a \cap \bx_{a'}$. 
 If intersection returns an ``empty" region or a region that we have already ``discovered" then it can be discarded.
We then further populate $C$ with the intersection of all pairs of child regions to obtain grandchildren and all pairs of grandchildren to obtain great-grandchildren. The procedure terminates when the set of regions at a given level is empty, so that no more intersections can be performed. We denote $U=P\cup C$, where $C$ is all of the regions discovered via intersection, as the total set of regions: these are the regions on which we will obtain approximate ``beliefs" from the GBP algorithm.  

We assign a counting number or ``Moebius" number to each region $r$ via the inclusion-exclusion principle
\begin{equation}
\sum_{r' \in U, r' \supseteq r}c_{r'} = 1 \ \forall r\label{Eq:moebius}
\end{equation}
i.e., for a given region $r$, the sum of the counting numbers for the regions which are supersets of $r$ (including $r$ itself) is one. We can compute Moebius numbers recursively, starting with the largest regions (i.e. the parents) for which $c_{r} = 1  \ \forall r \in P$.

The approximation of the free energy based on our choice of regions is given by 
\begin{equation}
\cF_{\rm GBP}(p)= \sum_{r \in U}c_r \sum_{\bx_r} p_r(\bx_r)\log \frac{p_r(\bx_r)}{F_r(\bx_r)}\label{eq:Fgbp}
\end{equation}
where we define the region factor / tensor
\begin{equation}
F_r(\bx_r)=\prod_{v\subseteq r} t_v(\bx_v),
\label{eq:factortensor}
\end{equation}
via the Hadamard product of all tensors for which $\bx_v\subseteq \bx_r$. If there are no tensors for which $\bx_v\subseteq \bx_r$ then we have $F_r(\bx_r) = 1 \ \forall \ \bx_r$.

It is easy to verify that $-\sum_{r\in U} c_r \log F_r(\bx_r)=-\sum_v \log t_v(\bx_v)$ and thus exactly gives the expected energy term in ~\autoref{eq:F}. The GBP approximation is made in the entropy term $-\sum_\bx p(\bx)\log p(\bx)\approx -\sum_{r\in U} c_r\sum_{\bx_r}p_r(\bx_r)\log p_r(\bx_r)$. Since the Moebius numbers can be negative, the entropy approximation is a difference of convex functions, making this a non-convex optimization problem. The quantity $\cF_{\rm GBP}$ is often referred to as the ``Kikuchi'' free energy.

\subsection{Message-passing equations}\label{sec:messagepassing}
We must optimize $\cF_{\rm GBP}$ with respect to the beliefs $p_r$ subject to the constraint that overlapping beliefs agree upon marginalization to the indices for which they intersect. For this it is sufficient to enforce the constraints between parent and child regions only~\cite{NIPS2001}:
\begin{equation}
p_a(\bx_b)\equiv \sum_{\bx_{a} \setminus \bx_{b}} p_a(\bx_a) = p_b(\bx_b)\quad b\in C, a\in P(b)\label{eq:constraints}
\end{equation}
Here we define $C(a)=\{b\in C|b\subset a\}$ as the set of child regions of region $a\in P$ and $P(b) = \{a\in P|a\supset b\}$ as the set of parent regions of region $b \in C$.

To solve the constrained optimization problem, we introduce Lagrange multipliers $\lambda_{ab}(\bx_b)$ and add the term $\sum_{b\in C}\sum_{a\in P(b)} \sum_{\bx_b} \lambda_{ab}(\bx_b)(p_a(\bx_b)-p_b(\bx_b))$ to Eq.~\ref{eq:Fgbp}. Minimizing $\cF_{GBP}$ with respect to $p_a(\bx_a)$ and $p_b(\bx_b)$ yields the solution in terms of the Lagrange multipliers, which we can write as
\bea
p_a(\bx_a) &=&\frac{1}{Z_a} F_a(\bx_a)\prod_{b \in C(a)} \m_{ab}(\bx_b)\quad a\in P \label{palpha1}\\
p_b(\bx_b) &=& \frac{1}{Z_b}F_b(\bx_b) \prod_{a  \in P(b)}\m^{-\frac{1}{c_{b}}}_{ab}(\bx_b)\quad b\in C\label{pbeta1}
\eea
where $\m_{ab}(\bx_b) =e^{\lambda_{ab}(\bx_b)}$ are the ``message tensors'' and $Z_a,Z_b$ ensure normalization of the beliefs, i.e. $\sum_{\mathbf{x}_{r}}p_{r}(\mathbf{x}_{r}) = 1$.

~\autoref{palpha1} and~\autoref{pbeta1} define the beliefs in terms of the messages $\m_{ab}$. We now must find the messages such that the beliefs satisfy the constraints
Eq.~\ref{eq:constraints}. An approach that is simple and works well in practice is to realize that for each $b \in C$  one can solve the constraint equations $p_a(\bx_b)=p_b(\bx_b)$  by changing $\m_{ab}(\bx_b)$ with $a\in P(b)$ simultaneously, while keeping all other messages constant.
This iteration scheme, first proposed in Ref.~\cite{NIPS2001} for factor graphs, reduces to the standard message passing scheme for simple BP.

The resulting update equations are~\cite{NIPS2001}
\begin{equation}
\m_{ab}(\bx_b)^\text{new}= \left(F_{b}(\mathbf{x}_{b})\right)^{d_{b}}\prod_{a' \in P(b)}
\left(
\frac{p_{a'}(\bx_b)}{\m_{a'b}(\bx_b)}\right)^{A_{a,a'}}
\label{gbp}
\end{equation}
where we have defined the numbers
\begin{align}
&d_{b} = \frac{c_{b}}{c_{b} + \vert P(b) \vert} \\
&A_{a,a'} = \frac{1}{c_{b} + \vert P(b) \vert} - \delta_{a,a'},
\end{align}
and $F_{b}(\bx_{b})$ and $p_{a'}(\bx_b)$ are given by Eqs.~\ref{eq:factortensor}
, \ref{eq:constraints}, and \ref{palpha1}.

The algorithm is initialized by a suitable choice of $\m_{ab}(\bx_b)$; for instance, $\m_{ab}(\bx_b)=1$ is often (though not always~\footnote{A simple example where uniform initialization of all messages is \textit{not} a good choice is in the ferromagnetic phase of the Ising model, where this uniform initialization is an unstable ``paramagnetic'' fixed point of the update equations.}) a reasonable initialization.

{\color{black}For non-negative tensors, these updates can also be carried out in the log domain. With $\lambda_{ab}=\log (\m_{ab})$,~\autoref{gbp} becomes
\begin{multline}
\lambda_{ab}^{\rm new}(\bx_b) = d_b\log F_b(\bx_b)\\
+\sum_{a'\in P(b)} A_{a,a'}\bigl[\log p_{a'}(\bx_b)-\lambda_{a'b}(\bx_b)\bigr],
\label{eq:gbp-log}
\end{multline}
up to an additive normalization constant.}

\subsection{Extracting observables}
After convergence, the GBP solution satisfies
\begin{equation}
\cF_{\rm GBP}(p)=-\sum_{r \in U}c_r\log \left( Z_{r} \right).
\end{equation}
and so the GBP approximation to the free energy (i.e. the negative log of the contraction of the tensor network) can be read off from the local partition functions $Z_r$.

A common goal with tensor networks is to contract the network down to a subset of tensors $t = \{t_{v_{1}}, t_{v_{2}} \hdots t_{v_{m}} \}$ with total indices $\bx_A=\left(\cup_{i=1}^{m}\mathbf{x}_{v_{i}} \right)$. In general, 
$\bx_A=(\bx_i, \bx_e)$ is the union of a set of boundary indices $\bx_e$ (the Markov blanket) and its interior $\bx_i$.
Therefore we are interested in calculating 

\begin{equation}
\frac{\partial Z}{\partial t_{v_{1}}\partial t_{v_{2}}\hdots \partial t_{v_{m}}} = \sum_{\bx \setminus \mathbf{x}_{e}}\frac{T(\bx)}{\prod_{i =1}^{m}t_{v_{i}}(\mathbf{x}_{v_{i}})}.
\end{equation}
Note, that this quantity does not depend on $\bx_i$. If the set $\bx_A$ is a subset of the indices $\bx_{r}$ for a region $r$ (either parent or child) then this quantity can be approximated directly from the GBP solution as
\begin{equation}
\frac{\partial Z}{\partial t_{v_{1}}\partial t_{v_{2}}\hdots \partial t_{v_{m}}}  \approx \sum_{\mathbf{x}_{r} \setminus \mathbf{x}_{e}} \frac{p_{r}(\mathbf{x}_{r})}{\prod_{i =1}^{m}t_{v_{i}}(\mathbf{x}_{v_{i}})}.
\label{eq:tnderivative}
\end{equation}
Such derivatives are required to measure local expectation values or optimize the local tensors in a tensor network.

If we wish to calculate a derivative with respect to a set of tensors whose combined indices $\bx_A=\left(\cup_{i=1}^{m}\mathbf{x}_{v_{i}} \right)$ are not a subset of any specific region (parent or child) we need to form an approximate belief on that region $p(\bx_A)$, given the parent/child beliefs that we do have. This can be done by stitching together the obtained beliefs of all parent/child regions $r$ {\color{black}contained within $A$}, i.e

\begin{equation}
p_{A}(\bx_A)=\prod_{r {\color{black}\subseteq} A} p_r(\bx_r)^{c_r'}.
\label{Eq:Blanket}
\end{equation}
Here, the counting numbers $c'_{r}$ are determined by applying the exclusion-inclusion principle to these GBP sub regions $r$, ie. setting $U=A$ in~\autoref{Eq:moebius}. One can then use the obtained belief $p_{A}(\bx_A)$ in \autoref{eq:tnderivative}. In the case of simple BP this derivative can be shown to reduce to the outer product of the messages from the indices $\mathbf{x}_{e}$ to the parent regions (i.e. individual tensors) that define the boundary of $t$.

\subsection{Recovering Simple Belief Propagation}
How does the element-wise message update equation ~\autoref{gbp} recover the simple BP update equations for a tensor network, which can just be written in terms of tensor contractions? Simple BP corresponds to taking a choice of parent regions consisting of one region for the indices associated with each tensor, ie. $P=\{\bx_1,\bx_2,\ldots,\bx_N\}$ with $\bx_a$ the indices of tensor $t_a$ and  $a=1,\ldots,N$.  The child regions are formed from parent intersections $b = (a \cap a')$.
For a tensor network, they are single bond indices $x_b=\bx_a\cap \bx_{a'}$ that are shared by adjacent tensors $a, a'$. As these child regions don't overlap, these are the only child regions and so $C$ is simply the set of all single bond indices. The factors are then $F_a(\bx_a)=t_a(\bx_a), a\in P$ and $F_b(\bx_b)=1, b\in C$. 
Since each bond $b$ appears in exactly two tensors, the Moebius numbers are $c_{a} = 1$, $c_b=-1$ giving $A_{a,a'} = 1 - \delta_{a,a'}$ with $P(b) =  \{a, a' \}$. Substituting ~\autoref{palpha1} into ~\autoref{gbp} gives
\begin{equation}
\m^\text{new}_{ab}(\bx_b) =\sum_{\bx_{a} \setminus \mathbf{x}_{b}}t_{a}(\bx_a) \prod_{b'\in C(a), b'\ne b}\m_{ab'}(x_{b'}),
\label{Eq: SimpleBP}
\end{equation}
which we recognize as the standard BP message passing equations.
Note that the product is over individual non-intersecting bond indices, allowing us to distribute the sum over the terms in the product. It can therefore be summed over ``on-the-fly", allowing one to write 
\begin{align}
&\m^\text{new}_{ab} = t_{a} \star \left(\star_{\substack{b'\in C(a) \\ b'\ne b}} \m_{ab'} \right),
\label{Eq: SimpleBP1}
\end{align}
with the star product meaning tensor contraction. For GBP, in general, however, this tensor contraction formulation does not immediately work because i) different messages may overlap so that element-wise products must be done before sums are taken and ii) children can have more than two parents.

\par In~\autoref{fig:GBPRegions} we illustrate a small square lattice tensor network and two choices of different GBP parent regions.  We show the simple BP case and an example of two generalized cases we consider in this work, which correspond to adding parent regions that capture the smallest plaquettes of the tensor network. These are the ``minimal'' region choices in order to go beyond BP and find fixed point message tensors that capture the fundamental loop correlations.

\par {\color{black} It is also interesting to note that the block BP algorithm introduced in Ref.~\cite{PhysRevB.108.125111} is recovered in our GBP framework when we partition the tensors into blocks and take all the indices of each block as one parent region. The child regions then consist of the indices shared by neighboring blocks, and the Kikuchi free energy reduces to the Bethe free energy on the blocked network. Meanwhile, Tensor network message passing (TNMP), introduced in Ref.~\cite{Wang2024TNMP}, uses overlapping neighborhoods and describes their surroundings with cavity messages~\cite{Newman2019,Kirkley2021,hack2025}. Its updates follow from local contractions without invoking a region-based free energy. In GBP instead, the updates follow from the principle of demanding stationarity of the Kikuchi free energy.}

\subsection{Positivity}
The GBP problem formulated in~\autoref{eq:F}, subject to the overlap constraints~\autoref{eq:constraints} for tensor networks differs in an important way from the traditional Kikuchi free energy or generalized belief propagation setting. In the traditional setting, the factors of the factor graph are entry-wise positive and so beliefs $p_r(\bx_r)$ are optimized over the simplex (non-negative and normalized) which is a compact domain. For positive factors, continuity on this compact feasible domain ensures that the free energy has a minimum, although it may lie on the boundary rather than satisfy the interior Euler--Lagrange equations. In the case of a tensor network, however, the local factors can contain negative or complex entries, so the domain may not be compact, which can dramatically affect the performance and convergence of GBP as we will see.
\par Specifically, where the tensor network $Z$ is the norm network for a tensor network state $\vert \psi \rangle$, we have $p(\bz,\bz')=\frac{1}{Z}\prod_v t_v(\bz_v,\bz_v')$ with $t_v(\bz_v,\bz_v')$ a positive semi-definite matrix that is not necessarily entry-wise positive. The messages $\m_{ab}(\mathbf{x}_{b}) = \m_{ab}(\mathbf{z}_{b}, \mathbf{z}'_{b})$ and beliefs can also be interpreted as matrices. Here, the choice of simple BP regions is very appealing because the update equation~\autoref{Eq: SimpleBP} preserves positive-semi-definiteness of the messages and consequently beliefs. Thus, if the messages are initialized as such, they will remain so and $\cF_{\rm GBP}$ in~\autoref{eq:Fgbp} is real, bounded from below and a proper objective which can be optimized.
\par A more general choice of regions, however, leads to a message update step requiring raising elements to non-trivial powers and cannot be written in terms of just Hadamard products of matrices. As a result it is not psd-preserving and can lead to non-psd beliefs and a non-real, unbounded free energy. {\color{black}For real signed networks, we evaluate negative arguments on a fixed side of the branch cut.} In~\autoref{fig:RandomNormNetworks}, we will see that a small number of negative elements in the norm network still lead to convergent, accurate GBP results, while beyond a threshold the algorithm fails to converge. We will meanwhile show results for networks such as the partition function of the frustrated classical Ising model or the norm of the AKLT state where the elements of the tensors are always positive and GBP performs very favorably.

\begin{table}

\label{Complexity}
\noindent \textbf{Flat Tensor Network} \\
\begin{tabularx}{\columnwidth}{@{\extracolsep{\fill}} |c c c c|}  
\hline
 Lattice & BP & \makecell{Cluster Expansion \\ (Order $n=1$)} & \makecell{GBP \\ (R1 Plaquettes)}  \\
 \hline
 Square & $\mathcal{O}(\chi^{4})$ & $\mathcal{O}(\chi^{4})$ & $\mathcal{O}(\chi^{4})$ \\
 Hexagonal & $\mathcal{O}(\chi^{3})$ & $\mathcal{O}(\chi^{3})$ & $\mathcal{O}(\chi^{3})$ \\
 Triangular & $\mathcal{O}(\chi^{6})$ & $\mathcal{O}(\chi^{6})$ & $\mathcal{O}(\chi^{6})$ \\
 Cubic & $\mathcal{O}(\chi^{6})$ & $\mathcal{O}(\chi^{6})$ & $\mathcal{O}(\chi^{6})$ \\
 \hline
\end{tabularx}

\vspace{1em} 

\noindent \textbf{Norm Network} \\
\begin{tabularx}{\columnwidth}{@{\extracolsep{\fill}} |c c c c|} 
 \hline
 Lattice & BP & \makecell{Cluster Expansion \\ (Order $n=1$)} & \makecell{GBP \\ (R1 Plaquettes)} \\
 \hline
 Square & $\mathcal{O}(\chi^{5})$ & $\mathcal{O}(\chi^{6})$ & $\mathcal{O}(\chi^{7})$ \\
 Hexagonal & $\mathcal{O}(\chi^{4})$ & $\mathcal{O}(\chi^{6})$ & $\mathcal{O}(\chi^{6})$ \\
 Triangular & $\mathcal{O}(\chi^{7})$ & $\mathcal{O}(\chi^{7})$ & $\mathcal{O}(\chi^{10})$ \\
 Cubic & $\mathcal{O}(\chi^{7})$ & $\mathcal{O}(\chi^{7})$ & $\mathcal{O}(\chi^{10})$ \\
 \hline
\end{tabularx}

\caption{Computational scaling with bond dimension of different message-passing based contraction algorithms for two different tensor network types and several different lattices. Algorithms are simple BP, a 1st order cluster expansion around the BP fixed point and GBP with R$1$ plaquette regions where there is a parent region for the indices associated to each tensor in the network and a parent region for the internal indices in each of the fundamental plaquettes of the network. 
Top table corresponds to a ``flat'' tensor network with a single layer of tensors (such as a classical partition function) and virtual indices of dimension $\chi$. Bottom table is for a ``double-layer'' network such as the norm $\langle \psi \vert \psi \rangle$ for the tensor network state $\vert \psi \rangle$ which has virtual indices of dimension $\chi$ and local physical dimension $d \ll \chi$. The scaling of these algorithms with system size $n$ is always linear for a fixed number of iterations of the corresponding message passing scheme.}
\label{table:complexity}
\end{table}

{\color{black}\subsection{Choosing Regions}\label{sec:regions}
The choice of parent regions is a key practical decision when applying GBP. A natural starting point is to include the smallest loops of the tensor network, allowing GBP to capture correlations which are missed by simple BP. On regular lattices these are the fundamental plaquettes. In three dimensions one can also consider regions formed from closed cages of plaquettes, as we do for the ice models. For the irregular planar networks considered in this work, we use the faces of the graph in the same way.

\par From these loops or cages one can construct the R$1$ and R$2$ choices we illustrate in~\autoref{fig:TN}. For R$1$, we take parent regions consisting of the internal indices of each loop or cage, alongside separate parent regions consisting of all indices incident to each tensor. For R$2$, we extend each loop or cage region to include all indices incident on its vertices. The child regions are then obtained from their intersections.

\par For a more general network, where there is no obvious choice of plaquettes or cages, one can instead identify short cycles or small groups of densely connected tensors. Our implementation, using the \texttt{TensorNetworkQuantumSimulator.jl} library \cite{rudolph2025simulatingsamplingquantumcircuits} accepts user-supplied parent regions or automatically constructs them from all simple cycles up to a user-set length. These provide candidate regions whose size must be balanced against the cost of storing and updating the corresponding beliefs and messages. It is typically best to start with the shortest loops and consider larger regions where computational resources permit. Larger regions allow more correlations to be treated and typically improve accuracy, but monotonic improvement is not guaranteed, and using larger regions can also make convergence more difficult.

}\subsection{Computational Complexity}
Before moving on to specific examples of our GBP implementation, it is worth commenting on the complexity, in terms of dimension of the bonds in the tensor network, of performing message tensor updates within the GBP paradigm. This is highly dependent on the order in which one performs the summations and the products in the message update equation and so it is much harder to identify an optimal complexity when the regions go beyond those of simple BP. Nonetheless, for a few lattices and the choice of R$1$ plaquette regions we have derived a relatively efficient message update cost in terms of the bond dimension $\chi$. We state them in~\autoref{table:complexity} alongside simple BP and a first order cluster expansion \cite{gray2025tensornetworkloopcluster, evenbly2025loopseriesexpansionstensor} for comparison. 
It is notable that for flat networks such as partition functions, GBP message updates are no more expensive than simple BP and yet the messages, if converged, will generally be more accurate due to capturing more complex correlations. For norm networks GBP with R$1$ regions induces a higher message update cost than simple BP, with the increased complexity of the local factors and messages allowing less room for optimization over a naïve treatment where the local double factors are first contracted together. Nonetheless, we find careful choice of the order in which each of the indices are treated in~\autoref{gbp} allows one to bring the scaling down from this naive $\mathcal{O}(\chi^{2z})$ to $\mathcal{O}(\chi^{z + \lceil \frac{z}{2} \rceil + 1})$ on the lattices considered --- where $z$ is the co-ordination number. We emphasize that this is not necessarily the optimal scaling, just a clear improvement over the most naïve approach.

\begin{figure*}[t]
    \centering
    \includegraphics[width=\textwidth]{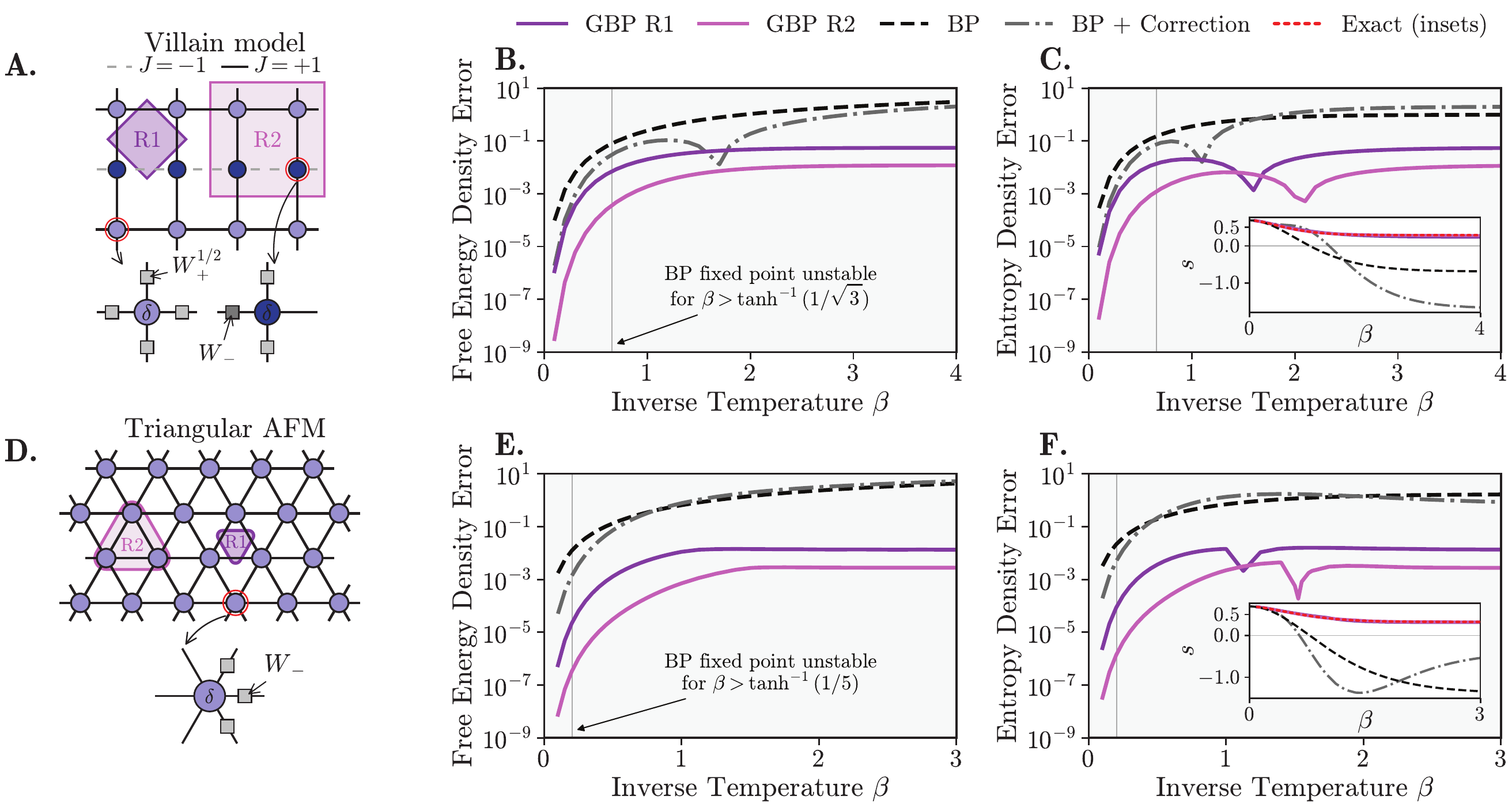}
    \caption{{\color{black}GBP for the Villain model (\textbf{A--C}) and triangular lattice antiferromagnet (\textbf{D--F}).
    \textbf{A, D.} Tensor network representations of the partition functions, with all indices of dimension $2$. Shading illustrates R$1$ (purple) and R$2$ (pink) regions. The local tensors are formed from copy tensors $\delta$ and Boltzmann matrices $W_{\pm}=\begin{psmallmatrix}e^{\pm\beta}&e^{\mp\beta}\\e^{\mp\beta}&e^{\pm\beta}\end{psmallmatrix}$. In A, the two colors distinguish the two sublattice tensors; in D, each tensor absorbs three matrices $W_-$.
    \textbf{B, E.} Absolute error on the free energy density versus inverse temperature, compared to the exact solutions in~\autoref{Eq:Fexact} and Refs.~\cite{Houtappel1950, Wannier1950, Wannier1973}, respectively.
    \textbf{C, F.} Absolute error on the entropy density. Insets compare the entropy densities with the exact results (red dotted). The loop correction uses the elementary square and triangular loops, respectively. The BP curves follow the paramagnetic fixed points. Vertical lines mark the onset of BP instability under simultaneous updates at $\beta=\tanh^{-1}(1/\sqrt{3})$ for Villain and $\beta=\tanh^{-1}(1/5)$ for the triangular antiferromagnet. Cusps occur when the approximate value crosses the exact value.}}
    \label{fig:ClassicalVillainTN}
    \label{fig:TriangularAFM}
\end{figure*}

\section{Numerical and Analytical Results}
{\color{black}We now demonstrate GBP on a range of classical and quantum models, including frustrated Ising models, independent set counting on irregular graphs, the deformed AKLT quantum state and the norm of random tensor network quantum states. These examples highlight GBP's improved treatment of frustration and symmetry, its successful use on irregular geometries, and its convergence difficulties for signed tensors.} We use the regions illustrated in~\autoref{fig:TN} and~\autoref{fig:GBPRegions}, along with their higher-dimensional generalizations.

{\color{black}For the classical models, counting problems and the AKLT norm network, we use the log-domain updates in~\autoref{eq:gbp-log}, treating exact forbidden configurations as fixed zeros. Where convergence is difficult (typically when multiple stationary points emerge), we supplement these with double-loop majorization of the Kikuchi free energy~\cite{yuille02,hes03a}. The outer loop replaces the concave entropy terms ($c_r<0$) by their linearization at the current feasible beliefs. The inner loop minimizes the resulting convex upper bound subject to normalization and the overlap constraints in~\autoref{eq:constraints}, using GBP updates with modified counting numbers and potentials. The outer iteration targets the original GBP stationary equations and, with exact inner minimization, yields a monotonically non-increasing Kikuchi free energy.

{\color{black}For the random norm networks, we use~\autoref{gbp} with linear damping on the messages and fix a branch convention when taking fractional powers of negative numbers. Damping and update order affect convergence but leave the fixed-point equations unchanged. No double loop fallback is used and implementing it for non-positive tensors is an open question.}

{\color{black}For all examples, we check convergence using the maximum disagreement between parent and child beliefs,}
\begin{equation}
\mathcal{R}=\max_{b\in C,\,a\in P(b)}\sum_{\bx_b}\left|p_a(\bx_b)-p_b(\bx_b)\right|,
\label{Eq:Convergence}
\end{equation}
with normalized beliefs and a tolerance of $10^{-7}$ ($10^{-8}$ for the counting results {\color{black}and $10^{-10}$ for the cubic random networks}).

More numerical details are provided in ~\appref{app:convergence}, including the specific models and the parameter regions where we need to invoke a double loop update to stabilize the GBP updates and find a stationary solution.}

\subsection{Classical Villain Model}\label{sect:villain}
As our first example we consider the partition function of the fully frustrated Ising model, also known as the Villain model, on a square lattice with Hamiltonian~\cite{Villain1977}
\begin{equation}
H = -\sum_{ \langle i,j \rangle}J_{i,j}s_{i}s_{j},
\end{equation}
where $J_{i,j} = - 1$ if $i$ and $j$ are both in an odd row of the lattice and $J_{i,j} = +1$ otherwise. This choice of couplings makes each plaquette frustrated. Unlike the unfrustrated Ising model, this model is known to lack a finite-temperature phase transition: it is disordered at all finite temperatures, while at $T=0$ it has quasi-long-range-order (QLRO) and an extensive entropy arising from an exponential ground-state degeneracy~\cite{Fisher1963,Fisher1963a,Forgacs1980}.

It is straightforward to construct a ``flat'' tensor network representation of the partition function $Z = \sum_{s_{1},s_{2}, \hdots}\exp(- \beta H)$ with bond dimension $2$ on any given square lattice, with the elements of each tensor all positive. We use the tensor construction shown in~\autoref{fig:ClassicalVillainTN}A. We consider three different choices of parent regions: i) one for each tensor (\autoref{fig:TN} Simple BP), ii) one for each tensor and one for each of the internal indices for the plaquettes of size $4$ (\autoref{fig:TN} GBP R$1$) or iii) one for each of the internal and external indices for the plaquettes of size $4$ (\autoref{fig:TN} GBP R$2$). We can compare the resulting Kikuchi free energy we obtain with the  recently derived \cite{CARAVELLI2022}, analytical expression for the free energy density in the thermodynamic limit 
\begin{widetext}
\begin{equation}
f(\beta)
= -\lim_{N \rightarrow \infty} \frac{1}{N}\log(Z(\beta)) = -\left[
\ln\!\bigl(2\cosh(\beta)\bigr)
+\frac{1}{4\pi^{2}}\int_{0}^{\pi}\!d\phi\int_{0}^{\pi}\!d\theta\;
\ln\!\Bigl((1+z^{2})^{2}-2z^{2}(\cos\phi+\cos\theta)\Bigr)
\right],
\label{Eq:Fexact}
\end{equation}
\end{widetext}
where $z=\tanh(\beta)$. We also compare the entropy density $s(\beta) = \beta e(\beta) - f (\beta)$, where $e(\beta) = \lim_{N\rightarrow \infty}\frac{1}{N}\langle H \rangle$ is the energy density.

~\autoref{fig:ClassicalVillainTN}B--C shows our numerical results. The frustration in the lattice quickly renders the only simple BP fixed point unstable for $\beta > \tanh^{-1}(\frac{1}{\sqrt{3}})$ (see~\appref{app:villain}). Even if one evaluates the free energy and corresponding entropy at the unstable fixed point --- which corresponds to $\m_{ab}(\mathbf{x}_{b}) = 1$ --- the result becomes increasingly poor as the temperature is lowered and a cluster correction  \cite{gray2025tensornetworkloopcluster, midha2025beyond} does not remedy this.
{\color{black}Including the smallest loops as GBP regions, however, captures the extensive ground state entropy of the Villain model. At $\beta=4$ we find an entropy per site of $0.238$ for R$1$ and $0.280$ for R$2$ (\autoref{fig:ClassicalVillainTN}C, inset), both close to the exact ground state value $s_0=G/\pi\simeq0.2916$, where $G$ is Catalan's constant~\cite{Fisher1963}. In~\appref{app:villain} we instead leave the Boltzmann weights on the bonds and take each square of four spins and four bond interactions as a plaquette region. For this choice we can solve the GBP equations analytically and obtain $s_0^{\rm GBP}=\ln(3\sqrt{3}/4)\simeq0.2616$.}

{\color{black}

\subsection{Triangular Lattice Antiferromagnet}\label{sect:triangular}
As a second frustrated example we consider the antiferromagnetic Ising model $H = \sum_{\langle i,j \rangle} s_i s_j$ on the triangular lattice. Every triangle of the lattice is frustrated and, like the Villain model, the system is disordered at all temperatures with an extensive ground state entropy, $s_0 = 0.323066...$~\cite{Wannier1950, Wannier1973}, whilst its free energy is known exactly~\cite{Houtappel1950}. We absorb each antiferromagnetic Boltzmann matrix into one endpoint, giving a tensor with six indices at every site. We use the triangular faces to choose R$1$ and R$2$ regions, as shown in~\autoref{fig:TriangularAFM}D.

\par In \autoref{fig:TriangularAFM}E--F we see a similar picture to the Villain model. The simple BP fixed point gives the entropy density in the thermodynamic limit
\begin{equation}
s_{\rm BP}(\beta)=\ln (2)+3\ln(\cosh(\beta))-3\beta\tanh(\beta),
\end{equation}
which becomes negative below $T \approx 1.284$. The fixed point is unstable at low temperatures, and a loop correction fails to remedy its unphysical entropy. GBP, meanwhile, gives stationary solutions across the temperatures shown, with the R$2$ regions improving upon R$1$ by over an order of magnitude at high temperatures. At $\beta = 3$, GBP gives $s \approx 0.310$ and $s \approx 0.321$ with R$1$ and R$2$ regions respectively, the latter within $1\%$ of the exact value. For R$2$, different initializations can give distinct low-temperature solutions. For consistency, we retain the stationary solution with the lowest Kikuchi free energy found at each temperature (\appref{app:convergence}).
}

\subsection{Residual Entropy of Ice-type Models}\label{sect:ice}

\begin{table*}

\label{IceModels}
\begin{tabular*}{\textwidth}{@{\extracolsep{\fill}} |c c c c c c c c|} 
 \hline
 Model & BP & \makecell{Loop Series \\ ($n=6$)} & \makecell{Loop Series \\ ($n=14$)} & \makecell{GBP \\ (R1 Plaq.)} & \makecell{GBP \\ (R1 Vox.)} & \makecell{GBP \\ (R2 Vox.)} & Literature \\ [0.5ex] 
 \hline
 Cubic Diamond $I_{c}$ & 1.5 & 1.504115 & 1.506362 & 1.503989 & 1.505193 & 1.506653 & 1.507467(4) \\ \hline
 Hexagonal $I_{h}$     & 1.5 & 1.504115 & 1.506375 & 1.504263 & 1.505377 & 1.506603 & 1.507466(4) \\ [1ex] 
 \hline
\end{tabular*}

\caption{Exponential of the residual entropy density $e^{s_{0}}$ (see \autoref{Eq:ResidualEnt}) of ice models to six decimal places as computed by contracting the corresponding three-dimensional infinite tensor network with BP; series expansion with loops of up to order 6 and 14~\cite{Nagle1966}; 
and GBP with various region choices of increasing complexity: R1 Plaquettes corresponds to choosing a parent region for each tensor and for the six internal indices in each of the fundamental plaquettes of the lattice; R1 Voxels correspond to choosing parent regions for the indices of each tensor and for the internal indices in each of the fundamental voxels of the lattice; R2 Voxels correspond to choosing parent regions for the internal and external indices in each of the fundamental voxels. These voxels are the 10-vertex adamantane cages ($[6^4]$) of the cubic diamond lattice and the 12-vertex $[6^5]$ cages of the hexagonal lattice; for the latter we do not include the 8-vertex $[6^3]$ cages which, together with the $[6^5]$ cages, tile the hexagonal ice lattice. Current best-estimated Monte Carlo values from literature are included in the last column along with their uncertainties \cite{Kolafa2014}. We note that, for $I_{h}$, the MC value is in very close agreement with a recent CTMRG-based calculation \cite{CTMRGIce1}.
}
\label{table:icemodels}
\end{table*}

Next, we approximate the residual entropy of ice via tensor network contraction. The ice crystal is a lattice of oxygen atoms, where each atom is connected to 4 neighboring oxygen atoms by hydrogen bonds. The hydrogen atoms can be in two stable positions either closer to one or the other adjacent oxygen atom. Not all $2^4$ hydrogen atom positions around a single oxygen atom are allowed: they follow the so-called ``two-in-two out'' rule, or ice rule, with two hydrogen atoms at the close position and two at the far position~\cite{Pauling35,Slater1941}. The residual ground state entropy is the logarithm of the total number of such hydrogen configurations in the lattice.

To evaluate this entropy, we define a bond dimension $2$ tensor network $T$ on a regular lattice of $N$ sites and coordination number 4 with the tensor elements set by the two-in-two-out rule 
\begin{equation} \label{eq:ice-rule}
t_{v}(\mathbf{x}_{v}) = \begin{cases}
1 \ \sum_{i}x_{v,i} = 2 \\
0 \ {\rm otherwise},
\end{cases}
\end{equation}
with $x_{v,i} \in \{0, 1 \}$ encoding the position of the hydrogen atoms adjacent to oxygen atom $v$. Note that for coordination number 4, there are exactly 6 configurations around each vertex which satisfy the rule, hence the alternative name of ``six-vertex model''~\cite{baxter2007exactly}.
The residual entropy in the thermodynamic limit is then given by 
\begin{equation}
s_{0} = \lim_{N \rightarrow \infty}\frac{1}{N}\ln(Z)
\label{Eq:ResidualEnt}
\end{equation}
with $Z$ given by~\autoref{Eq:TensorNetwork}.
We can approximate this quantity directly with both BP and GBP by running the message updates until convergence and computing the Kikuchi free energy.

In~\autoref{table:icemodels} we tabulate our results comparing to known exact and approximate values in literature. A simple counting argument provided by Pauling~\cite{Pauling35} estimates $e^{s_0}=\frac{3}{2}$ independent of the specific lattice structure (only the coordination number is important). This value is reproduced by BP as it makes the same intrinsic approximation.
\par For the three dimensional hexagonal ice and cubic ice models, which correspond to a lattice of vertically stacked hexagonal layers and a diamond cubic lattice respectively, the most precise values are currently given by Monte Carlo and CTMRG-based PEPS simulations \cite{Hayashi_2021, Kolafa2014, CTMRGIce1}. Due to uncertainties still present in these methods, the question of whether the exact entropies coincide is still open --- although the latest estimates put them to be very close, with the deviation, at most, in the 5th decimal place. We quote the most accurate Monte-Carlo based estimates in the table and also provide high order (up to 14th order) series expansions as originally computed by Nagle \cite{Nagle1966}~\footnote{More precisely, the series expansion up to order $n$ quoted in~\autoref{table:icemodels} sums over loop corrections to the BP fixed point from Eulerian cycles containing up to $n$ degree-2 vertices and any number of degree-4 (``crossover'') vertices. Loops containing odd-degree vertices have weight zero~\cite{Nagle1966}.}.

The GBP estimates approach the latest Monte-Carlo values $e^{s_{0}} \approx 1.50747...$ with increasing size of the parent regions. Our first GBP approximation is based on R$1$ regions using the smallest plaquettes (made up of six internal indices) in the $3$D hexagonal and diamond cubic lattices. For this region choice we are also able to analytically solve the GBP equations (see~\appref{app:ice}), in agreement with our numerics. For the diamond lattice and R$1$ plaquettes we find
\begin{equation}\label{eq:s0-diamond}
e^{s_{0}}_{\rm GBP} = \frac{3}{2} \left(\frac{(1+c_p^5)^4(1-c_p^5)^8}{(1+c_p^6)^{10}}\right) = 1.503989...
\end{equation}
with $c_p$ a solution of the fixed point equation
\begin{equation}\label{eq:fp-diamond}
c_p = -\frac{3 c_p^5 + 1}{3 + c_p^5}.
\end{equation}

By instead moving to parent regions consisting of voxels (the smallest three-dimensional volumes in the lattices) we are able to significantly improve this result. Specifically, in the case where the parent region for a voxel involves all of internal indices and the boundary indices of the voxel we are able to obtain $e^{s_{0}} = 1.506653$ and $e^{s_{0}} = 1.506603$ for diamond and hexagonal ice respectively. These values, being within $0.06 \%$ of the current Monte Carlo estimates, are more accurate than going up to the $14$th order in a series expansion \cite{Nagle1966},  though it should be noted that the ``bare'' estimate from a truncated series can be improved upon by extrapolation~\footnote{We further note that Nagle's series involves disconnected loops, which are problematic for convergence; while a cluster or cluster cumulant expansion might be preferred, the loose convergence guarantee from Ref.~\cite{midha2025beyond} cannot be immediately applied since dense loops containing many degree-4 vertices do not decay sufficiently fast.}. The GBP values for hexagonal and cubic ice differ only in the fifth decimal place, consistent with the latest estimates that the two residual entropies are extremely close; as this difference is well below the GBP approximation error, however, it does not by itself determine whether the exact values coincide. To handle the substantial size of the parent region in the R$2$ voxel case ({\color{black}$28$ indices for diamond ice and $33$ for hexagonal ice}) we adopted a sparse-tensor implementation of GBP where we only kept track of non-zero entries in the beliefs and messages. For {\color{black}diamond ice}, this leads to a substantial memory saving with the largest belief tensor only possessing $\sim 14 \times 10^{3}$ non-zero values of a possible total of $\sim 27 \times 10^{7}$.

{\color{black}We identify the voxels automatically from the graph, as the closed surfaces formed from at most five of its hexagonal faces. For the diamond lattice these are the adamantane cages, which tile the whole space. For hexagonal ice the tiling contains cages of eight and of twelve vertices. Our results above use only the twelve-vertex cages.}

{\color{black}\begin{figure}[t]\color{black}
    \centering
    \includegraphics[width=\columnwidth]{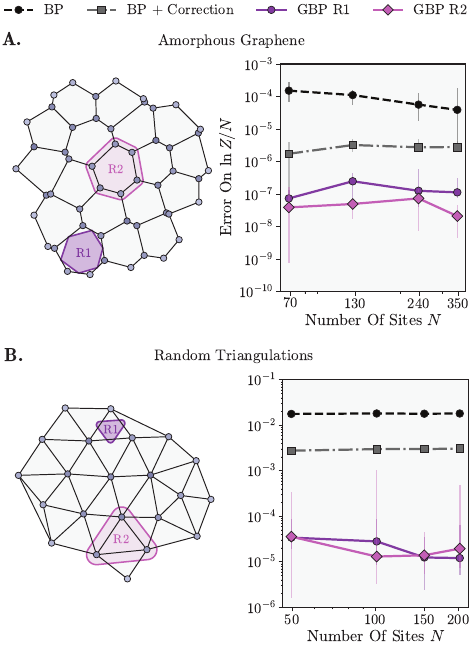}
    \caption{Independent-set counting on amorphous graphene flakes (\textbf{A}) and random triangulations (\textbf{B}). Each row shows a sample graph with R$1$ (purple) and R$2$ (pink) regions and the absolute error per site $|\ln Z-\ln Z_{\rm exact}|/N$. Colors distinguish the local tensors. We use 20 graphs of each type. Markers give the median and bars the range over five instances at each size. The loop correction includes loops of length up to eight in A and five in B. Walltimes are shown in~\autoref{fig:timings}.}
    \label{fig:HardCore}
\end{figure}}

{\color{black}

\subsection{Counting Independent Sets}\label{sect:hardcore}
We next count independent sets on irregular graphs. An independent set of a graph is a set of vertices no two of which are adjacent, and counting them is $\#$P-hard in general~\cite{Weitz2006}. For a graph $G=(V,E)$, the number of independent sets is the partition function of the hard-core lattice gas at unit activity,
\begin{equation}
Z = \sum_{\mathbf{n}\in\{0,1\}^{V}} \prod_{\{i,j\}\in E} (1-n_i n_j),
\label{eq:independent-sets}
\end{equation}
where $n_i$ is the occupation of vertex $i$ and the product runs over all edges of the graph, excluding configurations with adjacent occupied vertices. We write $Z$ as a tensor network in the same way as the Ising models above, with each vertex tensor copying its occupation number onto its incident bonds and the matrix $\begin{psmallmatrix} 1 & 1 \\ 1 & 0 \end{psmallmatrix}$ placed on each bond. In all cases we take the faces of the graph as GBP regions: R$1$ uses the face loops alongside the tensors and R$2$ uses each face along with all the indices incident on its vertices.

\par We consider two families of disordered planar graphs: flakes of amorphous graphene~\cite{Kumar2012Amorphous}, made up predominantly of rings of five, six and seven atoms, and random Delaunay triangulations. The former vary the sizes of the loops, while the latter vary the number of neighbors at each vertex, testing GBP beyond regular lattices. We construct both from uniformly distributed random points at unit density in a square, applying 15 steps of Lloyd relaxation: each point is moved to the centroid of its Voronoi cell~\cite{Du1999}. For the graphene flakes we retain the edges and vertices of cells whose generating points lie within a disk. For the triangulations we instead triangulate the points inside the disk, discarding triangles with any edge of length $2.2$ or greater to remove long boundary connections.

\par For these we obtain exact results by variable elimination for graphs of up to 351 and 212 sites respectively, and compare the methods in~\autoref{fig:HardCore}. For amorphous graphene (panel A), GBP gives median errors per site of $10^{-7}$ (R$1$) and $3 \times 10^{-8}$ (R$2$), roughly three orders of magnitude below BP and over an order below the loop correction. On the triangulations (panel B), R$1$ reduces the median error from $2 \times 10^{-2}$ to $2 \times 10^{-5}$. R$2$ has a similar median but larger worst-case errors, so larger regions need not always be more accurate.

\par Walltime comparisons are given in~\appref{app:convergence}, where GBP performs very favorably against the approximate model counter ApproxMC~\cite{Pote2025ApproxMC}, demonstrating consistently low errors and impressive walltimes across a wide range of system sizes.

}

\subsection{Deformed AKLT State on the honeycomb lattice}\label{sect:aklt}
Next, we move on to the quantum setting. We consider contracting the norm of the tensor network for the ``deformed" spin 3/2 AKLT quantum state  $\vert \psi(a) \rangle$ on the honeycomb lattice where $a$ is the deformation parameter \cite{PhysRevB.94.165130, PhysRevB.98.014432}. Specifically, the state is defined as 
\begin{equation}
\vert \psi(a) \rangle = \left(\otimes_{i=1}^{N}D_{i}(a)\right) \vert \psi_{\rm AKLT} \rangle
\label{Eq:AKLT}
\end{equation}
with $D_{i}(a)$ the local non-unitary deformation matrix $D_{i}(a) = {\rm diag}(\frac{a}{\sqrt{3}}, 1, 1, \frac{a}{\sqrt{3}})$ such that $a = \sqrt{3}$ directly corresponds to the AKLT state. The state $\vert \psi(a) \rangle$ is the unique ground state of the parent Hamiltonian
\begin{equation}
H = \sum_{\langle i, j \rangle} D^{-1}_{i}(a)D^{-1}_{j}(a) h_{i,j} D^{-1}_{i}(a)D^{-1}_{j}(a)
\end{equation} 
with
\begin{equation}
h_{i,j} = (\vec{S}_i \cdot \vec{S}_j) + \frac{116}{243} (\vec{S}_i \cdot \vec{S}_j)^2 + \frac{16}{243} (\vec{S}_i \cdot \vec{S}_j)^3.
\end{equation}
The deformation $a\ne \sqrt{3}$ breaks the SU(2) symmetry of the AKLT model to a $\mathbb{Z}_2$ symmetry around the $z$ direction and an O(2) symmetry in the xy plane. This allows for the occurrence of three different phases: an XY phase, the AKLT phase and a N\'eel phase as $a$ is varied.

The corresponding bond dimension $2$ honeycomb tensor network state can be constructed as an iTNS with the local $C_{3v}$ tensors $A^{s}_{i,j,k}$ and $B^{s}_{i,j,k}$ which have non-zero elements
\begin{align}\label{eq:aklt-peps}
A_{000}^{\Uparrow} &= a, \quad A_{100}^{\uparrow} = A_{010}^{\uparrow} = A_{001}^{\uparrow} = 1, \\ \notag
A_{111}^{\Downarrow} &= a, \quad A_{101}^{\downarrow} = A_{110}^{\downarrow} = A_{011}^{\downarrow} = 1 \\ \notag
B_{000}^{\Uparrow} &= a, \quad B_{101}^{\uparrow} = B_{110}^{\uparrow} = B_{011}^{\uparrow} = -1, \\ \notag
B_{111}^{\Downarrow} &= -a, \quad B_{100}^{\downarrow} = B_{010}^{\downarrow} = B_{001}^{\downarrow} = 1.
\end{align}
and reside on the honeycomb A and B sublattices respectively with $s \in \{\Uparrow, \uparrow, \downarrow, \Downarrow \}$ running over the four basis states for spin-$3/2$ and $i,j,k \in \{0,1 \}$.

\begin{figure}[t]
    \centering
    \includegraphics[width=\columnwidth]{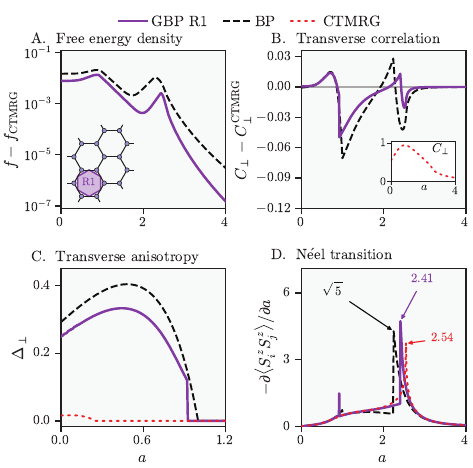}
    \caption{{\color{black}Deformed AKLT state on the honeycomb lattice. \textbf{A.} Free energy density difference $f-f_{\rm CTMRG}$, with $R=128$ for CTMRG. The inset shows the norm network and an R$1$ plaquette. \textbf{B.} Difference $C_\perp-C_\perp^{\rm CTMRG}$ in the transverse correlation $C_\perp=-\tfrac12\langle S_i^x S_j^x+S_i^y S_j^y\rangle$ for nearest neighbors $i,j$. The inset shows $C_\perp$ from CTMRG. \textbf{C.} Transverse anisotropy $\Delta_\perp=\tfrac12|\langle S_i^x S_j^x\rangle-\langle S_i^y S_j^y\rangle|$ at small $a$ which should be zero. CTMRG retains a small anisotropy at the lowest $a$. \textbf{D.} Derivative of the longitudinal correlation. Arrows mark the predicted N\'eel transitions. The known critical point is $a_c\simeq 2.54$~\cite{PhysRevB.94.165130}.}}
    \label{fig:AKLT}
\end{figure}

Here we compare results from contracting the norm of this infinite tensor network state $\langle \psi(a) \vert \psi(a) \rangle$ and measuring local correlators $\langle S^{\alpha}_{i}S^{\alpha}_{j} \rangle$ with $\alpha \in \{x,{\color{black}y},z\}$. We note that the entries of the norm tensors are all non-negative, despite the negative entries in the state itself, with the norm being directly mappable to the partition function of a classical eight-vertex model \cite{PhysRevB.94.165130} for large $a$. We compare results from BP, GBP with R1 regions and {\color{black}reference results} from our implementation of CTMRG for infinite hexagonal $C_{3v}$ tensor networks \cite{HexCTMRG1, HexCTMRG2}. {\color{black}Analytical expressions for the BP fixed points and correlators are given in~\appref{app:aklt}. We use CTMRG at environment dimension $R=128$ as the reference throughout~\autoref{fig:AKLT}.} To compute the derivative of the tensor network with respect to neighboring pairs of tensors (which is necessary to evaluate nearest-neighbor observables) with GBP we follow the procedure described earlier: see~\autoref{eq:tnderivative} and \autoref{Eq:Blanket}. {\color{black}We cancel the local norm factors and repeated message factors algebraically before evaluating this environment, avoiding division by zero.}

\par In~\autoref{fig:AKLT} we see that GBP shows an improvement on the error in the free energy density \newline $f = -\lim_{N \rightarrow \infty}\frac{1}{N}\log \langle \psi(a) \vert \psi(a) \rangle$ compared to simple BP. This error is highest for small $a$ where, for $a \lesssim 0.42$ the system is known to be in an XY phase with quasi-long range order - making the environment highly correlated and requiring a high CTMRG environment dimension for accurate results. For intermediate $0.42 \lesssim a \leq 2.54$ this system is in an AKLT phase with topological properties and no local order. Meanwhile for $a \geq 2.54$ the system transitions into a Néel phase consisting of a superposition of two antiferromagnetic states. This second order transition is witnessed via a divergence in the derivative with respect to $a$ of the local ZZ order $\langle S^{z}_{i}S^{z}_{j} \rangle$ (\autoref{fig:AKLT}D). Both BP and GBP observe this transition but at different critical points, with the GBP transition occurring closer to the true transition point. 

{\color{black}At small $a$, BP and GBP give relatively accurate estimates of the transverse correlation $C_\perp$ (\autoref{fig:AKLT}B), with GBP offering a small improvement. Both methods however, show appreciable anisotropy (panel C) reflecting the fact they break the O(2) symmetry. GBP reduces this anisotropy, which disappears at the selected branch change near $a\simeq0.92$, compared with $a=1$ for BP. The same branch change produces the feature in panel D. At small $a$, both methods do admit O(2) symmetric fixed points, but these have higher free energy and are unstable to symmetry-breaking perturbations.}

{\color{black}At small $a$, the CTMRG free energy is converged after $1600$ sweeps at $R=128$, but the transverse magnetization is still changing slowly and thus there is a small breaking of the O(2) symmetry at that $R$.}

\subsection{Random Networks}\label{sect:random}

Finally, we consider a tensor network state $\vert \psi \rangle$ on a {\color{black}square or cubic lattice} where each tensor $\psi_{v}(\mathbf{z}_{v}, s_{v})$ possesses a single physical index $s_{v}$ of dimension two and a number of virtual indices $\mathbf{z}_{v}$. We draw the values of $\psi_{v}(\mathbf{z}_{v}, s_{v})$ uniformly, at random, via
\begin{equation}
\psi_{v}(\mathbf{z}_{v}, s_{v}) \sim U(- \alpha, 1 - \alpha) \qquad 0\le \alpha \le 1
\end{equation}
and then construct the norm tensor network with ``double factor" tensors $T_{v}(\mathbf{x}_{v}) = T_{v}(\mathbf{z}_{v}, \mathbf{z}_{v}') = \sum_{s_{v}}\psi_{v}(\mathbf{z}_{v}, s_{v})\psi^{*}_{v}(\mathbf{z}'_{v}, s_{v})$. 
For $\alpha = 0$ or $\alpha = 1$ the double factors are entry-wise positive, while for $\alpha = 0.5$ each off-diagonal entry is equally likely to be positive or negative. Thus $\alpha$ allows us to control the fraction of ``negativity" in the elements of the tensors in the network --- with $T_v$ being Hermitian and PSD for all $\alpha$.

\par
\begin{figure}[t]
    \centering
    \includegraphics[width=\columnwidth]{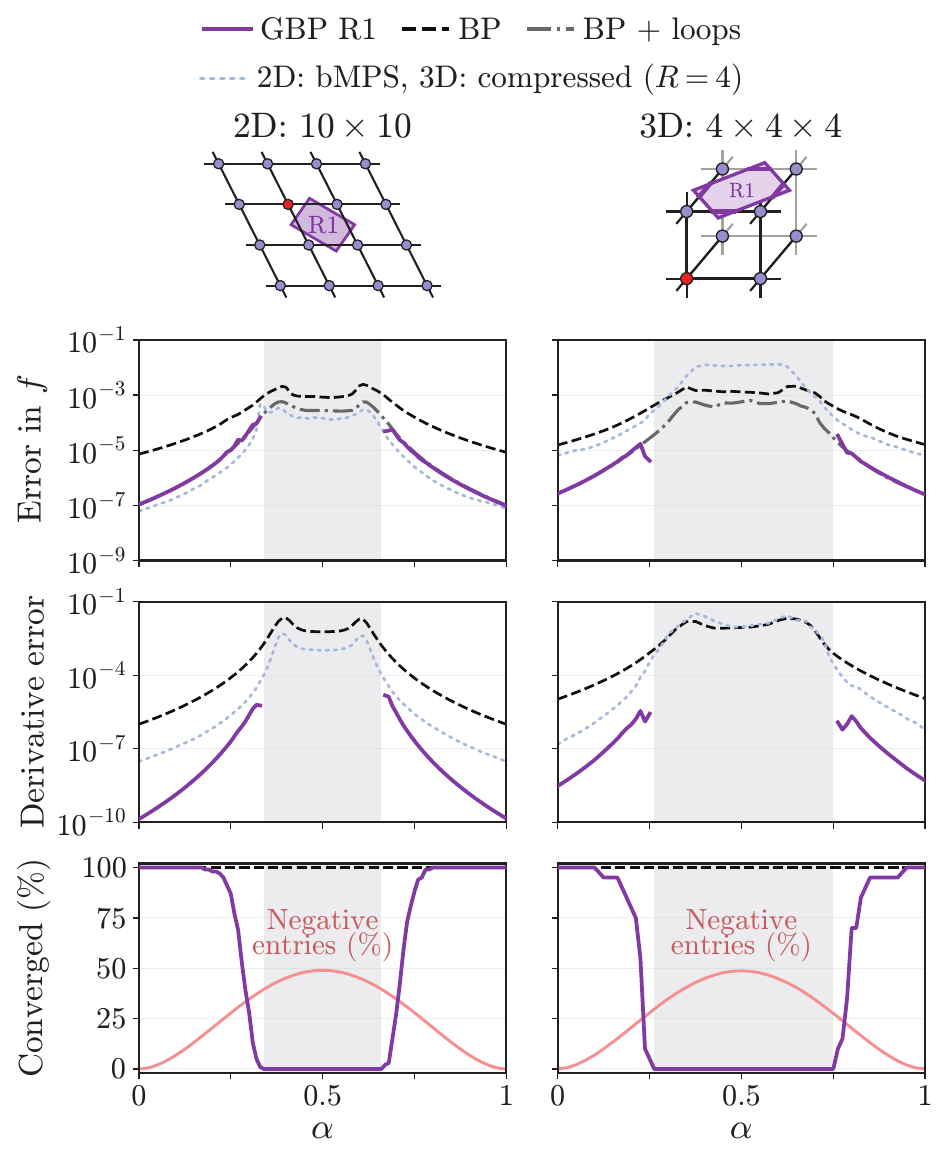}
    \caption{{\color{black}Random norm networks $\langle\psi\vert\psi\rangle$ with open boundaries. The left column shows $10\times10$ square lattices with $\chi=3$ ($100$ realizations per $\alpha$), and the right column shows $4\times4\times4$ cubic lattices with $\chi=2$ ($20$ realizations per $\alpha$). State-tensor entries are drawn from $U(-\alpha,1-\alpha)$. From top to bottom we show the absolute error in the free energy per site $f$, the squared-angle error $1-|\langle a,b\rangle|^2/(\lVert a\rVert^2\lVert b\rVert^2)$ in the derivative with respect to the central norm tensor, and the percentage of converged samples. The pink curve gives the percentage of negative tensor entries. The gray dash-dotted curve in the top row shows the first-order loop correction to BP. The blue curves show boundary MPS in 2D and compressed contractions~\cite{Gray2024} in 3D, both at $R=4$. See \appref{app:convergence} for the reference contractions. The diagrams above the plots show the network geometries and R$1$ regions. Errors are averaged over converged instances. Shading marks intervals with no converged GBP samples.}}
    \label{fig:RandomNormNetworks}
\end{figure}

 In the {\color{black}left column of~\autoref{fig:RandomNormNetworks}} we show results from running the GBP algorithm on a $10 \times 10$ square lattice and 100 independent realizations of this norm network. The state bond dimension is $\chi = 3$. We compute reference contractions and derivatives of this tensor network with respect to the centre double factor via boundary MPS with $R=128$.

\par For values of $\alpha \leq 0.2$ and $\alpha \geq 0.8$ we obtain accurate results under the GBP algorithm. {\color{black}At least $98\%$ of instances at each $\alpha$ in this range meet the convergence criterion in~\autoref{Eq:Convergence}}, yielding much improved results compared to BP for both the free energy density and especially for the derivative of the norm network around a central tensor~\autoref{eq:tnderivative}. These local environments are needed for measuring observables and updating individual tensors. In the case of the free energy we can also directly compute a cluster expansion around the BP fixed point which provides results of comparable accuracy on the free energy to GBP --- with both methods essentially capturing the first order ``loop" correlations. We note that it is possible, but less straightforward, to compute the derivative of the tensor network with a cluster expansion, whereas it is available directly from the GBP beliefs. We also highlight results from boundary MPS (bMPS) with a heavy level of truncation ({\color{black}$R=4$}). The error in bMPS improves rapidly with $R$ and drops below the GBP results for $R > 8$. 

We should emphasize we evaluate the bMPS free energy by substituting the converged MPS ``messages" into the formula for the Bethe free energy on a tree (where the nodes are column MPOs from the tensor network). This is far more accurate than just accounting for the loss of norm during truncation and contracting the right or left-most column with the incident, truncated MPS (which is the typical approach with bMPS \cite{PhysRevB.109.235102, Verstraete2004}). {\color{black}Timings for the different contraction approaches on representative 2D instances are shown in~\autoref{fig:timings}A.}

\par For intermediate values of $0.2 \leq \alpha \leq 0.8$ GBP with plaquette parents shows a sharp drop in convergence, {\color{black}with no instances converging for $0.34 \leq \alpha \leq 0.66$ in our scan.} Such an issue does not manifest in BP, which is aided by the psd-preserving nature of the update equations --- although we emphasize that this property does not guarantee convergence. Nonetheless, the non-convergent GBP regime occurs alongside a broad peak in the error of BP, BMPS and cluster corrections --- suggesting it is indicative of a broader complexity in contracting the given network for the range $\alpha$, potentially in the form of a transition into a ``hardness" regime. Such a sign-based hardness transition has also been observed in other tensor networks such as ``flat" networks with elements drawn uniformly at random or from Haar random distributions \cite{Gray2024, Chen2025}.

{\color{black}\par The right column shows the same calculation on a $4\times4\times4$ cubic lattice, using $\chi=2$ and $20$ realizations per $\alpha$. We again use the elementary square plaquettes as R$1$ regions. Here too, GBP dramatically improves the local environments where it converges, while loop corrections give comparable free-energy errors. We also compare with compressed contractions~\cite{Gray2024} with heavy compression at $R=4$, with timings in~\autoref{fig:timings}B. Again, the accuracy of this method improves rapidly with $R$ and becomes similar to GBP at $R = 8$.  The convergence difficulties for GBP and general increased errors in other methods persist over a broad interval around $\alpha=1/2$.}

\section{Conclusion}
{\color{black}In this work we have extended the generalized belief propagation (GBP) method to tensor network contraction and compared its accuracy on a range of classical and quantum problems. While the advantages of GBP over BP for classical problems are well established, our AKLT and random norm-network results demonstrate these advantages for quantum states.}

\par {\color{black}Across the examples considered, GBP typically gives accurate results where it converges, substantially improving on simple BP and, in several examples, loop corrections. For quantum states, this includes more accurate local environments for random norm networks and an improved estimate of the AKLT--N\'eel transition.} Particular success is observed in frustrated tensor networks, such as for the partition function{\color{black}s} of the fully frustrated Ising model {\color{black} and the triangular lattice antiferromagnet}, where strong local correlations which are resonant around the smallest loops in the network{\color{black},} lead simple belief propagation astray. {\color{black}GBP also gives accurate independent-set counts on disordered graphs.} {\color{black}Using a sparse-tensor implementation of GBP, we also approximate the exponential of the residual entropy density of three-dimensional ice structures to within $0.06\%$ of the latest Monte-Carlo estimates.} We anticipate future application of GBP to various tensor network contraction problems, including those representing the partition function of frustrated classical and quantum spin systems or those representing the solution to counting problems.

\par {\color{black}GBP's principal limitation is convergence, particularly for quantum norm networks containing negative or complex entries.} Unlike the simple BP update equations, the GBP update equations that we use here (and also other formalisms such as those of Yedidia \cite{BeliefPropagation4}) do not preserve positive semi-definiteness of the message tensors when contracting the norm of a quantum tensor network state $\langle \psi \vert \psi \rangle$. For non-negative networks, the double-loop method used here can recover stationary solutions when direct message updates are unstable~\cite{yuille02,hes03a}. For signed or complex beliefs, the convexity and real-valued free-energy bound used by this construction no longer apply. A potential route would be performing Riemannian gradient descent on the Kikuchi free energy within the manifold of PSD messages \cite{Lin2019, Bergmann2022}. {\color{black}Developing more reliable updates for these networks is therefore an important direction for future work.}

\section*{Acknowledgments}
The authors are grateful for ongoing support through the Flatiron Institute, a division of the Simons Foundation. They would also like to acknowledge Dries Sels and Matt Fishman for insightful discussions, and Siddhant Midha for collaboration on related work.
The tools and methods described in this paper were implemented with the \texttt{TensorNetworkQuantumSimulator.jl} library \cite{rudolph2025simulatingsamplingquantumcircuits}, an open-source Julia package built on top of ITensors.jl \cite{itensor-r0.3} for optimizing and contracting tensor networks of arbitrary topology.

\bibliography{bibliography}

\clearpage
\appendix
\begin{figure*}[ht]
\centering
\includegraphics[width=\linewidth]{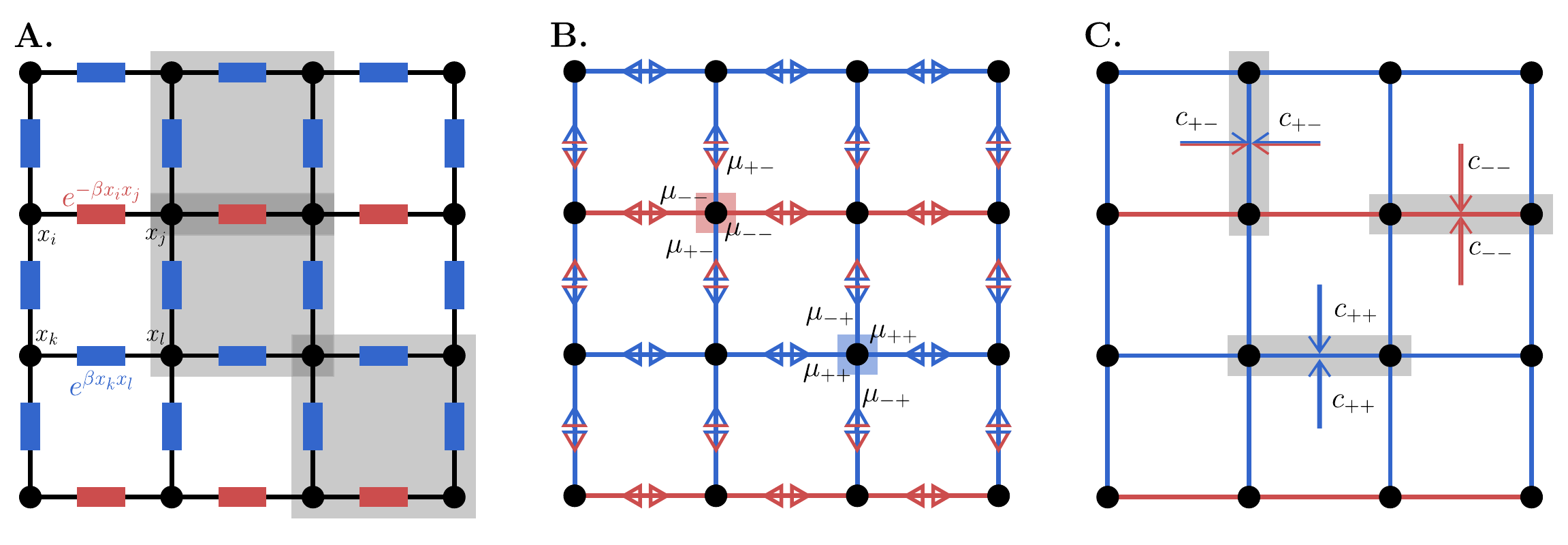}
\caption{\label{fig:villain-app} (Generalized) belief propagation on the classical Villain model. {\bf{A.}} Rewriting of the tensor network in~\autoref{fig:ClassicalVillainTN}A. Black vertices are the variables (delta tensors), blue rectangles are the ferromagnetic Boltzmann weights $e^{\beta x_i x_j}$, and red rectangles are the antiferromagnetic Boltzmann weights $e^{-\beta x_i x_j}$. The plaquettes shaded in gray are examples of parent regions for GBP, which intersect on both vertices and edges. {\bf{B.}} In simple BP, the child regions are vertices, which come in two types: $-$ vertices (such as the one shaded red) receive messages along horizontal antiferromagnetic bonds and vertical ferromagnetic bonds, parameterized by $\mu_{--}$ and $\mu_{+-}$ respectively, whereas $+$ vertices (such as the one shaded blue) receive messages along horizontal and vertical ferromagnetic bonds, parameterized by $\mu_{++}$ and $\mu_{-+}$ respectively. {\bf{C.}} In GBP with plaquette parent regions, the plaquette-edge messages come in three types: those directed towards vertical bonds (parameterized by $c_{+-}$), horizontal ferromagnetic bonds (parameterized by $c_{++}$), and horizontal antiferromagnetic bonds (parameterized by $c_{--}$).}
\end{figure*}
\section{Analytical solution to (G)BP Equations for Villain model}\label{app:villain}

In this Appendix we solve for fixed points of~\autoref{gbp} for the Villain model. {\color{black}Similar phenomenology is observed in the analytical solutions to (G)BP equations for the antiferromagnetic Ising model on the triangular lattice.}

In~\autoref{fig:ClassicalVillainTN}, we represented the partition function as a tensor network where each four-index tensor includes interactions between four pairs of spins. There we encounter the somewhat annoying feature that, in order to avoid introducing complex-valued entries that would stymie GBP, the full point group symmetry is not explicit at the level of the tensors. 

In deriving an analytical solution, it will be more convenient to leave the Boltzmann interactions on the bonds, so that each vertex tensor is simply a delta tensor (\autoref{fig:villain-app}A).
Then, passing to the factor graph language, we will consider two choices of parent regions: (1) each bond ``factor'' and its adjacent vertex ``variables,'' recovering simple BP, (2) each plaquette, which involves four bond factors and four vertex variables. We consider these in turn.

\subsection{Simple BP}
In simple BP, the messages are passed from edges to vertices. There are two types of vertices: those involved in four ferromagnetic interactions, which we will denote $+$ vertices, and those involved in two ferromagnetic and two antiferromagnetic interactions, denoted $-$. {\color{black}The four messages in~\autoref{fig:villain-app}B all have the form $(1+\mu,1-\mu)/2$. Horizontal ferromagnetic and antiferromagnetic bonds send messages with parameters $\mu_{++}$ and $\mu_{--}$, respectively, to either endpoint. A vertical bond sends a message with parameter $\mu_{-+}$ to its $+$ endpoint and $\mu_{+-}$ to its $-$ endpoint.}

The resulting update equations are:
\begin{subequations}
\begin{align}
\mu_{++}^{\text{new}} &= \tanh(\beta) f(\mu_{-+},\mu_{++}) \\
\mu_{-+}^{\text{new}} &= \tanh(\beta) f(\mu_{--},\mu_{+-}) \\
\mu_{+-}^{\text{new}} &= \tanh(\beta) f(\mu_{++},\mu_{-+}) \\
\mu_{--}^{\text{new}} &= -\tanh(\beta) f(\mu_{+-},\mu_{--})
\end{align}
\end{subequations}
where
\begin{equation}
f(x,y) = {\color{black}\tanh[2 \tanh^{-1}(x) + \tanh^{-1}(y)].}
\end{equation}

Clearly, $(\mu_{++}, \mu_{-+}, \mu_{+-}, \mu_{--})=(0,0,0,0)$ is a fixed point of these equations for all $\beta$. In fact, one can check (i.e., with the aid of symbolic computing software) that $(0,0,0,0)$ is the \textit{only} fixed point within the space of physical messages ($|\mu|\leq 1$). We call this fixed point ``paramagnetic'' because it respects the $\mathbb{Z}_2$ symmetry of the Hamiltonian: since the $\mu$'s are zero, the BP magnetization of each spin is also zero, and the BP belief $p_{ij}$ on edge $x_i,x_j$ is simply given by the local interaction term: $p_{ij}(x_i,x_j) \propto \exp(-\beta J_{ij} x_i x_j)$. 

There are two problems with this fixed point. First, it is completely insensitive to the frustration in the model: it is identical to the paramagnetic BP fixed point of the ferromagnetic Ising model, with an expected energy per edge of $-\tanh(\beta)$, or $-2\tanh(\beta)$ per spin. 
This yields qualitatively incorrect predictions as $\beta \rightarrow \infty$, most notably a ground state energy density that is much below the true ground state energy and a negative ground state entropy density. Second, this fixed point becomes unstable at $\tanh(\beta_c) = 1/\sqrt{3}$ 
, as can be verified by computing the eigenvalues of the linearized BP equations. Often, the instability of the paramagnetic fixed point coincides with the emergence of additional, symmetry-broken fixed points, signaling a transition (within the Bethe approximation) to an ordered phase. But in the Villain model, where $(0,0,0,0)$ remains the only fixed point at all temperatures, this means that there is no stable BP solution for $\beta > \beta_c$: the message passing algorithm initialized away from the unstable point will fail to converge. 

\subsection{GBP with plaquette regions}
Simple BP struggles with the Villain model because it does not account for the model's most salient feature: the frustration around each plaquette. In the main text, GBP R$1$ accounts for this frustration by including two types of parent regions, around vertices and plaquettes. In the factor graph language, the choice is even simpler: every plaquette is a parent region, and the children are edges and vertices. {\color{black}Although the resulting GBP equations are not identical to those implemented numerically for GBP R1 in the main text, both constructions capture the plaquette frustration in a minimal fashion. Our analytical results in this section therefore lend theoretical support to the empirical advantage of GBP over BP shown in~\autoref{fig:ClassicalVillainTN}.}

Here we focus on finding a fixed point with trivial single-variable beliefs. Note that all plaquette regions are identical, involving 3 positive edges ($J_{ij}=1$) and one negative edge ($J_{ij}=-1$). There are three types of plaquette-edge messages, targeting $++$, $+-$, and $--$ edges (see~\autoref{fig:villain-app}C). We parameterize these messages by 
\begin{equation}\label{eq:message-c}
\m_{p\rightarrow e}(x_1,x_2) = (1 + c_e x_1 x_2)/4
\end{equation}
with $e=(++),(+-),(--)$, respectively, so that $c_e$ measures the correlation along that edge. Further defining
\begin{equation}
g(x,\beta) = \frac{x + \tanh(\beta)}{1 + x \tanh(\beta)},
\end{equation}
we obtain, upon substitution of~\autoref{eq:message-c} into the message passing equation~\autoref{gbp}
\begin{subequations}\label{eq:gbp-villain}
\begin{align}
c_{++}^{\rm{new}} &= -g(-c_{--},\beta) g(c_{+-},\beta)^2, \\
c_{+-}^{\rm{new}} &= -g(c_{++},\beta) g(c_{+-},\beta) g(-c_{--},\beta), \\
c_{--}^{\rm{new}} &= g(c_{++},\beta) g(c_{+-},\beta)^2.
\end{align}
\end{subequations}

These equations admit a fixed point $(c_{++}, c_{+-}, c_{--}) = (c^*(\beta), c^*(\beta), -c^*(\beta))$, where 
$c^*(\beta) = -g(c^*(\beta),\beta)^3$. Linearizing~\autoref{eq:gbp-villain} around this fixed point shows that the solution remains stable, with all eigenvalues $|\lambda|<1$ at all temperatures (within the subspace of trivial single-variable beliefs). The time scale of convergence (determined by the inverse logarithm of the largest eigenvalue), however, diverges as $\beta \rightarrow\infty$. If one instead initializes the messages randomly (but in a translation-invariant manner), allowing for nontrivial single-variable beliefs, then we find that the fixed point is fully stable for $\beta \lesssim 1.092$ as long as a nonzero amount of damping is used, but acquires a weakly unstable direction for $\beta \gtrsim 1.092$.

The functional form of $c^*(\beta)$ is not particularly illuminating, so we do not present it here. More informative is the limiting behavior of the energy and entropy computed within the GBP approximation as $\beta \rightarrow \infty$. Unlike simple BP, GBP with plaquette parent regions correctly recovers the ground state energy density of $-1$ per spin. It also yields a much more accurate approximation of the ground state entropy density,
\begin{equation}
s_{\rm GBP}(T=0) = \log \left(\frac{3\sqrt{3}}{4} \right) = 0.2616\dots,
\end{equation}
a slight underestimate of the exact value, $G/\pi = 0.29156...$, where $G$ is Catalan's constant~\cite{Fisher1963}.


\section{Analytical solution to GBP Equations for Ice-type models}\label{app:ice}
In this Appendix we solve for fixed points of~\autoref{gbp} for spin-ice on the two lattices (hexagonal and diamond cubic) considered in the main text as well as on the square lattice, where the exact residual entropy is known \cite{Lieb1967}. We take the simplest case beyond simple BP, corresponding to R$1$ plaquette parent regions, where there are two types of parent regions, corresponding to the vertices and internal indices of the plaquettes of the tensor network (\autoref{fig:GBPRegions}). For the square lattice these plaquettes are loops of size $4$ whereas for hexagonal and diamond they are loops of size $6$. In the main text numerical results are given for more complicated parent region structures, which results in a higher accuracy on the residual entropy.

To derive analytical results, we focus once again on finding translation-invariant fixed points with trivial single-index beliefs, so as in~\autoref{eq:message-c}, the messages from parent regions to two-index intersections are parameterized by a single ``correlation parameter'' $c$. We denote this parameter $c_v$ ($c_p$) for messages from vertex (plaquette) parent regions, respectively. 

On the square lattice, the update equations reduce to
\begin{align}\label{eq:update-square}
c_v^{\rm{new}} = c_p^3, \quad c_p^{\rm{new}} = -\frac{3 c_v^2 + 1}{3 + c_v^2}.
\end{align}
Owing to the hard constraint (\autoref{eq:ice-rule}), the entropy density is simply the minus Kikuchi free energy density $s_0^\mathrm{sq}=-F(c_v,c_p)$.  Substituting $c_v = c_p^3$ yields
\begin{equation}\label{eq:c0-cp}
\exp\left[s_0^{(\mathrm{sq})}\right] = \frac{(1-c_p^3)^2(3 + 2 c_p^3 + 3 c_p^6)}{2(1+c_p^4)^3}.
\end{equation}
At the unique local minimum, which is a stable fixed point of~\autoref{eq:update-square}, we obtain $e^{s_0} = 1.518609...$, whereas the exact value is $\frac{8\sqrt{3}}{9} =1.539601...$.

On the diamond lattice, the update equations reduce to
\begin{subequations}
\begin{align}
c_p^{\rm{new}} &= -\frac{1+3c_v-c_p^5(3+c_v)}{3+c_v-{\color{black}c_p^5}(1 + 3 c_v)}\label{eq:cp-d} \\
c_v^{\rm{new}} &= \frac{2 c_p^5}{1 + c_p^{10}}.\label{eq:cv-d}
\end{align}
\end{subequations}
Substituting~\autoref{eq:cv-d} into~\autoref{eq:cp-d} yields the fixed point equation and entropy density in terms of a single variable quoted in the main text (\autoref{eq:fp-diamond} and \autoref{eq:s0-diamond}). 

The calculation for hexagonal ice is more involved, as now the ``internal index'' parent regions split into two types, and there are additional child regions supported on three indices. We simply quote the final result: $e^{s_0} = 1.5042627...$. 

\section{BP Equations for the Deformed AKLT Model}\label{app:aklt}
In this section we obtain the analytical BP solution for the deformed AKLT model. This analysis reproduces the simple BP numerical results of~\autoref{fig:AKLT} and provides insight into the violation of O(2) symmetry at small $a$. 

First, we obtain the ``double factors'' on the A and B sublattice of the norm network. Recall that a double factor is obtained by contracting a PEPS tensor with its adjoint along the physical degree of freedom, e.g.,
\begin{equation}
    t_A(\bz,\bz') = \sum_s A^s_{\bz} (A^*)^s_{\bz'}
\end{equation}
and likewise for $t_B$, where $s$ runs over the spin-3/2 basis states $\{\Uparrow,\uparrow,\downarrow,\Downarrow\}$. Substituting~\autoref{eq:aklt-peps}, we find that the double factors on the A and B sublattices are identical: $t_A = t_B \equiv t$. Labeling the incident edges $(z_1,z_1'),(z_2,z_2'),(z_3,z_3')$, with each index taking values in $\{0,1\}$,
\begin{widetext}
\begin{equation}\label{eq:t-aklt}
t(z_1,z_1',z_2,z_2',z_3,z_3') =
\begin{cases}
a^2 & \text{all indices 0 or all indices 1} \\
1 & \sum_i z_i = \sum_i z_i' = 1 \text{ or } \sum_i z_i = \sum_i z_i' = 2 \\
0 & \text{otherwise}.
\end{cases}
\end{equation}
\end{widetext}
Since the same double factor appears on each site, we seek solutions with identical messages $m_i(z_i,z_i')=m(z,z')$ on each edge $i=1,2,3$. The messages $\m(z,z')$ are psd Hermitian matrices, which we will take to be real symmetric and $\sum_{z,z'}m(z,z')=1$. Therefore, they can be parameterized by two parameters {\color{black}$(\mu,c)$}, so that
\begin{align}\label{eq:aklt-messages}
    \m(z,z') = \frac{1}{4}&\left((1+(-1)^z \mu)(1+(-1)^{z'} \mu)\right. \notag \\
    - &\left.c (-1)^{\delta_{z,z'}}\right).
\end{align}

Substituting the parameterization~\autoref{eq:aklt-messages} and the factors~\autoref{eq:t-aklt} into the BP update equation~\autoref{Eq: SimpleBP} yields a system of two equations for $(\mu^{\rm{new}},c^{\rm{new}})$ in terms of $(\mu,c)$. The fixed points come in three types, where we call a fixed point ``physical'' if the message is Hermitian and psd, and stable if both eigenvalues of the Jacobian matrix have absolute value $<1$:
\begin{enumerate}
    \item $\left(\mu=0, c = \frac{3 - a^2 - 2 \sqrt{2(1-a^2)}}{1 + a^2}\right)$ is stable for $a < 1$, but unphysical for $a > 1$.
    \item $(\mu=0,c=1)$ is a physical fixed point for all $a$, but is only stable in the interval $(1,\sqrt{5})$. On the endpoints of this interval, it is marginally stable, and coincides with fixed points (1) and (3) respectively.
    \item $\left(\mu=\pm \sqrt{\frac{a^2-5}{a^2-1}}, c=\frac{4}{a^2-1}\right)$ is unphysical for $a<\sqrt{5}$ and stable for $a>\sqrt{5}$.
\end{enumerate}

{As a result, BP has three stable fixed points, one for each phase:}
\comment{As a result, the BP algorithm (with an arbitrary initialization away from a fixed point) will converge to three different fixed points for the three phases:} fixed point (1) for $a<1$, fixed point (2) for $1<a<\sqrt{5}$, and fixed point (3) for $a>\sqrt{5}$. At the boundaries between these regimes, two fixed points merge and become marginally stable ($|\lambda_0|=1$) which means that the distance from the fixed point decays only algebraically with the number of BP iterations rather than exponentially.

By sandwiching the operator $S_x$, $S_y$, or $S_z$ in between the bra and ket and contracting BP messages along the virtual legs, we can obtain local expectation values. The one-point expectations satisfy $\langle\vec{S}_{i\in A}\rangle_{\rm BP} = - \langle \vec{S}_{i\in B} \rangle_{\rm BP}$, with:
\begin{subequations}
\begin{align}
   \langle S_{i\in A}^x\rangle_{\rm BP} &= 
   \begin{cases}
       \frac{3 (a^2-2 a\sqrt{3}-5) \sqrt{2(1-a^2)}}{8 (a^2-3)} & a < 1 \\
       0 & a \geq 1,
       \end{cases}
        \\
   \langle S_{i \in A}^y \rangle_{\rm{BP}} &= 0, \\
   \langle S_{i\in A}^z \rangle_{\rm{BP}} &=
   \begin{cases}
       0 & a \leq \sqrt{5} \\
       \pm \frac{3 \sqrt{(a^2-5)(a^2-1)}}{{\color{black}2}(a^2-3)} & a > \sqrt{5}. 
       \end{cases}
\end{align}
\end{subequations}
The nearest neighbor correlators in the three regimes are:
\begin{widetext}
\begin{subequations}\label{eq:correlators}
\begin{align}
    \langle S_i^x S_j^x \rangle_{\rm{BP}} &= \begin{cases} 
   -\frac{81 + 60 a \sqrt{3} - a^2 (38 + 44 a \sqrt{3} + 15 a^2)}{32 ({\color{black}3-a^2})} & a \leq 1 \\
   -\frac{4 + 4 \sqrt{3} a + 3 a^2}{(3 + a^2)^2} & 1 < a < \sqrt{5} \\
   -\frac{16 + a (a^2 - 3) (8 \sqrt{3} - 9 a + 3 a^3)}{2 (a^2 - 3) (a^2 - 1)^3} & a \geq \sqrt{5}
   \end{cases} \\
   \langle S_i^y S_j^y \rangle_{\rm{BP}} &= \begin{cases}
       -\frac{25 + a (20 \sqrt{3} + a (2 - 4 a\sqrt{3}+ a^2))}{32 (3 - a^2)} & a < 1 \\
    \langle S_i^x S_j^x \rangle_{\rm{BP}} & a \geq 1
   \end{cases} \\
   \langle S_i^z S_j^z \rangle_{\rm{BP}} &= \begin{cases}
   -\frac{(1 + a^2)^2}{8 (3 -a^2)} & a \leq 1 \\
   -\left(\frac{1+3a^2}{2(3 + a^2)}\right)^2 & 1 < a < \sqrt{5} \\
   -\frac{9 a^6 - 45a^4 + 31 a^2 - 27}{4 (a^2 - 3) (a^2-1)^2} & a \geq \sqrt{5}.
    \end{cases}
\end{align}
\end{subequations}
\end{widetext}
The BP approximation correctly captures certain qualitative and quantitative aspects of the phase diagram: it detects SSB of the $\mathbb{Z}_2$ symmetry at large $a$, recovers $\lim_{a\rightarrow\infty}\langle S_i^z S_j^z\rangle = -9/4$ and $\lim_{a\rightarrow\infty} \langle S_i^x S_j^x \rangle = \lim_{a\rightarrow\infty} \langle S_i^y S_j^y\rangle = 0$, and has the full SU($2$) symmetry (all correlators equal) at the undeformed AKLT point ($a=\sqrt{3}$). However, BP notably fails for $a<1$, as the stable fixed point breaks the XY symmetry, wrongly predicting a nonzero X magnetization and large discrepancy between the XX and YY correlators. 
{\color{black}
\section{Numerical Details, Convergence Analysis, and Walltimes}\label{app:convergence}
The update in~\autoref{gbp} can be performed in many different ways. In a simultaneous, or Jacobi, scheme every new message is computed from the messages of the previous sweep, whilst in a sequential scheme the children are visited in turn and all messages into a child are updated using the latest values of the others~\cite{NIPS2001}. Children sharing no parent can be updated in parallel. These schemes have the same fixed points but different stability properties.
Log-domain updates can also be used to help stability (see Eq.~\ref{eq:gbp-log}), although stability is not guaranteed.

\par Where direct GBP updates fail to converge for nonnegative tensors, we use double-loop majorization of the same Kikuchi free energy~\cite{yuille02,hes03a}. Write $c_b$ for a child's counting number and set $t_b=\max(c_b,0)$. Linearizing the concave entropy terms about the current fixed beliefs $\bar p$ gives a convex upper bound $Q(p \mid \bar p)$ to the free energy, satisfying
\begin{equation}
Q(p\mid\bar p)-\mathcal{F}(p)=\sum_b(t_b-c_b)D_{\rm KL}(p_b\Vert\bar p_b)\geq0.
\end{equation}
In the inner loop, we minimize $Q$ by sequentially updating the Lagrange multipliers enforcing parent--child consistency. The update equations are the same log-domain updates as in the main text [\autoref{eq:gbp-log}], but with $c_b$ replaced by $t_b$ and the child log potential $c_b \log F_b$ replaced by $c_b\log F_b+(t_b-c_b)\log\bar p_b$. We hold $\bar p$ fixed until the overlap constraints are satisfied, then use the resulting beliefs to form the next bound. The bound is tight at $\bar p$, so exact constrained minimization of $Q$ decreases the original free energy, though finite inner tolerances weaken this guarantee. We check inner consistency, changes in the beliefs, and the residual of the original GBP equations reconstructed from the messages. We use this double-loop majorization in the following cases in the main text:
\begin{itemize}
    \item \textit{Triangular antiferromagnet (\autoref{fig:TriangularAFM}D--F):} R$1$ at $\beta=1$ and $1.125$, and R$2$ at all sampled points with $1.375\leq\beta\leq3$.
    \item \textit{Ice (\autoref{table:icemodels}):} R$1$ plaquettes for cubic ice and R$2$ voxels for hexagonal ice, both at zero temperature.
    \item \textit{Deformed AKLT state (\autoref{fig:AKLT}):} the selected solutions at $a=0.13$--$0.395$, $2.38$, $2.400$--$2.435$, $2.445$ and $2.56$, on a grid with spacing $0.005$. We also run majorization from uniform initializations throughout $0\leq a\leq4$ to check for competing branches. Sweep counts are shown in~\autoref{fig:aklt-sweeps}.
    \item \textit{Independent-set counting (\autoref{fig:HardCore}):} R$2$ at unit activity, for the amorphous graphene flakes with $N=68$, $126$, $136$, $228$, $234$, $239$, $240$, $253$, $337$, $346$, $349$, $350$ and $351$ sites, and the triangulations with $N=47$, $96$, $150$, $151$ (two instances), $153$, $194$, $198$ and $206$ sites.
\end{itemize}

\begin{figure}[t]\color{black}
    \centering
    \includegraphics[width=\columnwidth]{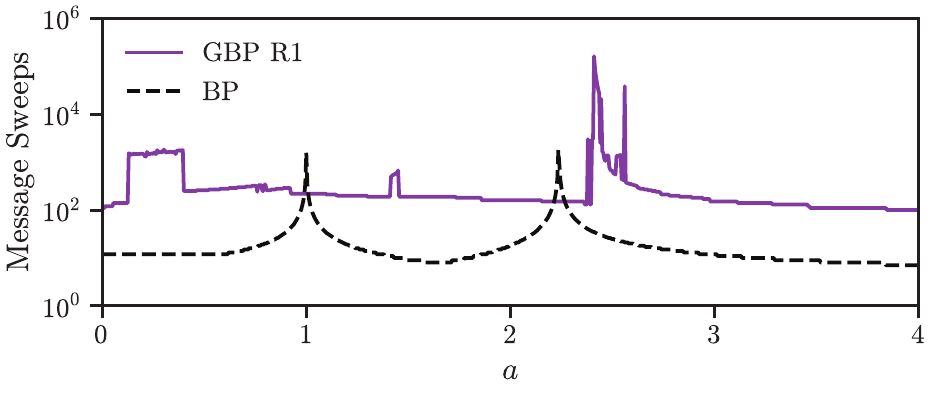}
    \caption{Message sweeps for BP and the selected GBP branch of the deformed AKLT state. GBP starts from a nearby converged solution. We count the inner majorization sweeps and omit additional branch checks.}
    \label{fig:aklt-sweeps}
\end{figure}

\par For the random networks, we use linearly damped GBP updates with weight $0.35$ and uniform initial messages: simultaneous updates with a limit of $3000$ sweeps in 2D, and sequential updates with a limit of $20000$ sweeps in 3D. Messages are normalized by their largest absolute entry to avoid underflow.

\par The 3D reference contractions use bond compression along optimized contraction paths~\cite{Gray2024}. We compare bond dimensions and independent paths, up to dimension $256$ for the norm and $128$ for the central environment.

\begin{figure*}[p]\color{black}
    \centering
    \includegraphics[width=\textwidth]{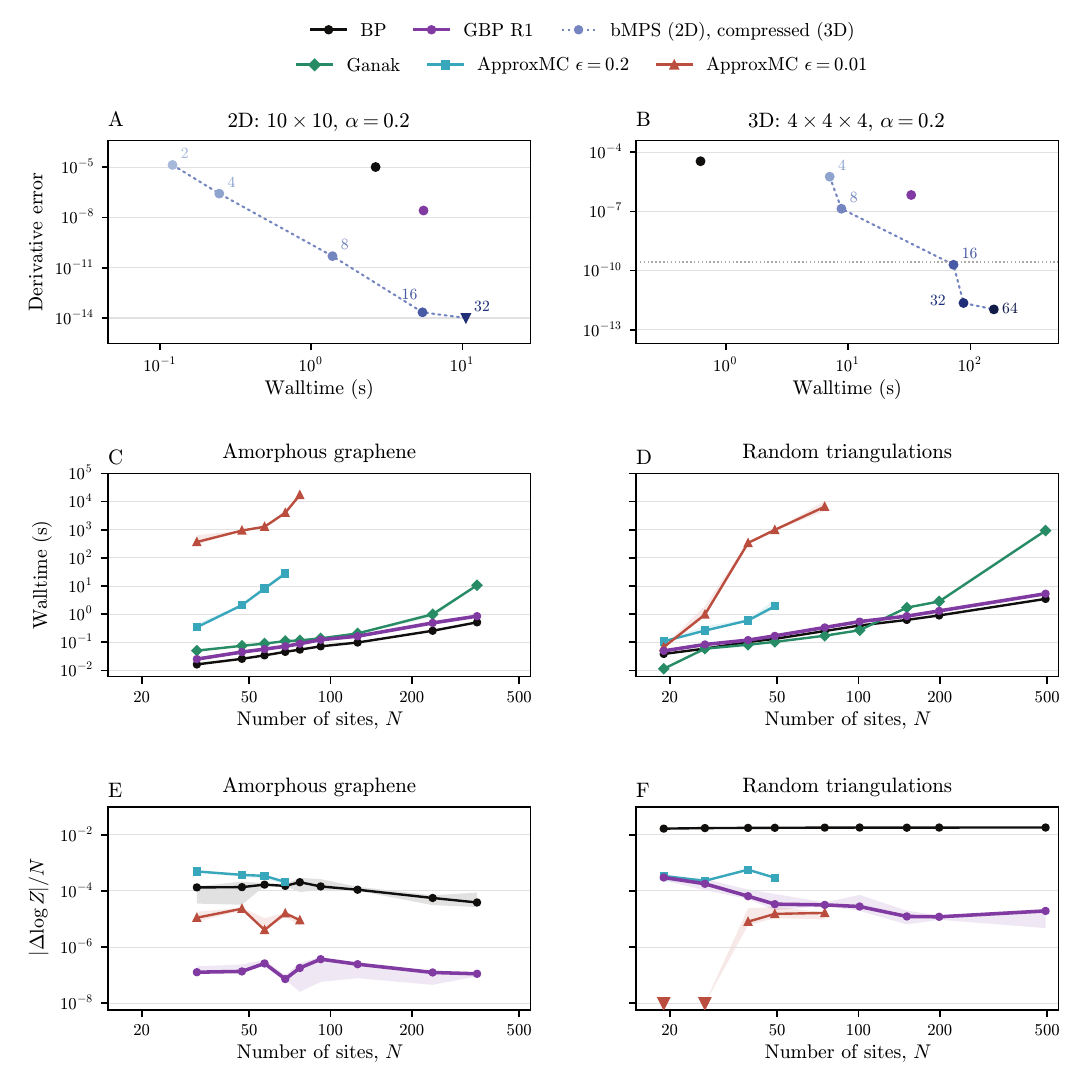}
    \caption{Timing comparisons. \textbf{A, B.} Derivative error versus walltime at $\alpha=0.2$. All methods are compared on the same instances, eight square networks in A and seven cubic networks in B. Errors and references are as in~\autoref{fig:RandomNormNetworks}. Blue labels give $R$. The grey line in B indicates the estimated accuracy of the 3D reference. At $\alpha=0.5$, GBP failed to converge on all eight timed instances in each dimension. \textbf{C--F.} Counting walltimes (C, D) and absolute errors per site in $\log Z$ (E, F) versus the number of sites. Reference counts use variable elimination or Ganak in exact mode. Ganak has zero error and is omitted from E and F. Points show medians across instances after taking the median of five repetitions each. Bands show the interquartile range across graphs. Counting results include completed runs only. Downward triangles denote errors below $10^{-14}$ in A and $10^{-8}$ in E and F, including zero. Panels C--F use an AMD EPYC 7742. All other timings use an Intel Xeon Platinum 8362.}
    \label{fig:timings}
\end{figure*}

\par We compare walltimes and errors in~\autoref{fig:timings}. The GBP, BP and Boundary MPS methods use TensorNetworkQuantumSimulator~\cite{rudolph2025simulatingsamplingquantumcircuits}, whilst compressed contractions use quimb and cotengra~\cite{Gray2024}. Timings depend strongly on the implementation. Within each comparison we timed all methods on the same processor model, using one core on Rusty after an untimed warm-up. Region construction, contraction-path optimization, iteration and observable evaluation are included where applicable. Input preparation and compilation are excluded.

At $\alpha=0.2$ on the random norm networks, R1 GBP's errors are comparable to boundary MPS or compressed contraction with bond dimensions between $4$ and $8$ but requiring around a quarter of the GBP walltime.

GBP fares best in a walltime vs error comparison on the irregular counting problems of~\autoref{sect:hardcore}. On the largest main-text triangulations, R$1$ took $1.3$ seconds in a separate comparison with the first-order loop correction (length-five cutoff), which took $15$ seconds. The R$1$ error was more than two orders of magnitude smaller. R$2$ is computationally much more expensive and improves the graphene results, but not the triangulations. In~\autoref{fig:timings}C--F we compare counting times and errors up to roughly $500$ sites, including the dedicated counters Ganak~\cite{Ganak2019} and ApproxMC~\cite{Pote2025ApproxMC}. GBP has a modest cost above BP, while remaining much more accurate where exact counts are available. GBP also converges on graphs up to $2125$ sites, beyond the available exact references. ApproxMC uses relative tolerances $\epsilon=0.2$ and $0.01$ with $95\%$ confidence, providing a probabilistic error guarantee which GBP does not have, but scaling much more severely in walltime. Ganak is exact, but appears substantially more costly around $N\sim500$.
}

\end{document}